\documentclass[a4paper,12pt]{article}
\usepackage{latexsym}
\usepackage{amssymb}
\usepackage{amsfonts} 
\usepackage{amsmath}
\usepackage[mathscr]{euscript}

\title{\textsf{The bipolaron in the strong coupling limit}}
\date{\empty}

\author{
Tadahiro Miyao\thanks{This work was supported  by Japan Society for the Promotion of
Science (JSPS).
Permanent address: The graduate school of natural science and technology,
Okayama university, Okayama 700-8530, Japan, e-mail:
\texttt{tmiyao@math.okayama-u.ac.jp} }
  and  Herbert Spohn\footnotemark[2]\\ 
{\it Zentrum Mathematik,}
{\it Technische Universit\"at M\"unchen,}\\ 
{\it  D-85747 Garching, Germany}\\
e-mail:
\footnotemark[1]\texttt{ miyao@ma.tum.de},
\footnotemark[2]\texttt{ spohn@ma.tum.de}  
}

\newcommand{\one}{{\mathchoice {\rm 1\mskip-4mu l} {\rm 1\mskip-4mu l}
{\rm 1\mskip-4.5mu l} {\rm 1\mskip-5mu l}}}
\newcommand{\h}{\mathfrak{h}}
\newcommand{\Hil}{\mathcal{H}}
\newcommand{\ex}{\mathrm{e}}
\newcommand{\D}{\mathrm{dom}}

\newcommand{\im}{\mathrm{i}}
\newcommand{\Fock}{\mathfrak{F}}
\newcommand{\Ffin}{\mathfrak{F}_{\mathrm{fin}}}

\newcommand{\dG}{\mathrm{d}\Gamma}

\newcommand{\me}{m_{\mathrm{e}}}

\newcommand{\hotimes}{\hat{\otimes}}

\newcommand{\la}{\langle}
\newcommand{\ra}{\rangle}

\newcommand{\wlim}{\mbox{$\mathrm{w}$-$\displaystyle\lim_{n\to\infty}$}}
\newcommand{\slim}{\mbox{$\mathrm{s}$-$\displaystyle\lim_{n\to\infty}$}}

\newcommand{\BbbR}{\mathbb{R}}
\newcommand{\BbbN}{\mathbb{N}}

\newcommand{\BbbC}{\mathbb{C}}
\newcommand{\vepsilon}{\varepsilon}
\newcommand{\vphi}{\varphi}
\newcommand{\Pt}{P_{\mathrm{tot}}}
\newcommand{\Pf}{P_{\mathrm{f}}}
\newcommand{\Nf}{N_{\mathrm{f}}}
\newcommand{\Hb}{H_{\mathrm{bp}}}
\newcommand{\Hbk}{H_{\mathrm{bp},\kappa}}

\newcommand{\Hsk}{H_{\mathrm{p},\kappa}}
\newcommand{\Hs}{H_{\mathrm{p}}}
\newcommand{\Ebk}{E_{\mathrm{bp},\kappa}}
\newcommand{\Esk}{E_{\mathrm{p},\kappa}}
\newcommand{\Eb}{E_{\mathrm{bp}}}
\newcommand{\Es}{E_{\mathrm{p}}}
\newcommand{\Uni}{U_{\kappa,K}}
\newcommand{\A}{A_{\kappa,K}}
\newcommand{\Hk}{H_{\kappa,K}^{\mathrm{bp}}}
\newcommand{\Hki}{H_{\infty,K}^{\mathrm{bp}}}
\newcommand{\SHk}{H_{\kappa,K}^{\mathrm{bp}}}
\newcommand{\FullHil}{L^2(\BbbR^6,\mathrm{d}x_1\otimes\mathrm{d}x_2)\otimes\Fock(L^2(\BbbR^3))}
\newcommand{\OneHil}{L^2(\BbbR^3,\mathrm{d}\xr)\otimes\Fock(L^2(\BbbR^3))}
\newcommand{\xr}{x_{\mathrm{r}}}
\newcommand{\xc}{x_{\mathrm{c}}}
\newcommand{\vth}{\vartheta}
\newcommand{\ep}{\mathcal{E}_{\mathrm{p}}}
\newcommand{\eb}{\mathcal{E}_{\mathrm{bp}}}
\newcommand{\uni}{\mathcal{U}}

\begin{document}

\newtheorem{define}{Definition}[section]
\newtheorem{Thm}[define]{Theorem}
\newtheorem{Prop}[define]{Proposition}
\newtheorem{lemm}[define]{Lemma}
\newtheorem{rem}[define]{Remark}
\newtheorem{assum}{Condition}
\newtheorem{example}{Example}
\newtheorem{coro}[define]{Corollary}

\maketitle

\begin{abstract}
The bipolaron are two electrons coupled to the elastic deformations of
 an ionic crystal. We study this system in the Fr\"{o}hlich
 approximation. If the Coulomb repulsion dominates, the lowest energy
 states are two well separated polarons. Otherwise the electrons form a
 bound pair. We prove the validity of the Pekar-Tomasevich energy
 functional in the strong coupling limit, yielding  estimates on the
 coupling parameters for which the binding energy is strictly
 positive. Under
 the condition of a strictly positive binding energy we prove the
 existence of a ground state at fixed total momentum $P$, provided $P$
 is not too large.
\end{abstract}

\section{Introduction}
The polaron is an electron coupled to the elastic deformations of an
ionic crystal. We rely here on  the approximation proposed by
H. Fr\"ohlich \cite{HFroehlich}, where the phonons are represented  as a Bose field over
$\BbbR^3$, the dispersion relation is constant, $\omega(k)=\omega_0$, and
the coupling function is proportional  to $1/|k|$ in wave number
space. The Hilbert space of the polaron is then
$\Hil=L^2(\BbbR^3)\otimes\Fock(L^2(\BbbR^3))$ with $\Fock(L^2(\BbbR^3))$
the bosonic Fock space and the hamiltonian  is given by
\begin{align*}
\Hs
=-\frac{1}{2}\Delta_x\otimes\one+\sqrt{\alpha}\lambda_0
 \int_{\BbbR^3}\frac{\mathrm{d}k}{(2\pi)^{3/2}|k|}
\Big[\ex^{\im k\cdot x}\otimes a(k)+\ex^{-\im k\cdot x}\otimes
 a(k)^*\Big]+\one\otimes \Nf
\end{align*}
with $\lambda_0=(2\sqrt{2}\pi)^{1/2}$.
Here $\Delta_x$ is the Laplacian, $\Nf$ is the number operator, and
$a(k), a(k)^*$ are the bosonic annihilation and creation operators with
commutation relations 
\[
[a(k), a(k')^*]=\delta(k-k'),\ \ [a(k), a(k')]=0=[a(k)^*, a(k')^*]. 
\] 
(The complete definition of $\Hs$ will be recalled in the subsequent
section.)
We use units in which $\hbar=1, \omega_0=1,$ and the bare mass of the
electron $\me =1$. Since the coupling  function is pure power, the only
parameter in the model is the dimensionless coupling constant
$\sqrt{\alpha}$.

The bipolaron, the subject of our paper, consists of {\it two}
electrons coupled to the elastic deformations of an ionic crystal. The
Hilbert space  is then $\Hil=L^2(\BbbR^6)\otimes\Fock(L^2(\BbbR^3))$
and, in the Fr\"ohlich approximation, the hamiltonian  reads
\begin{align*}
\Hb=&\sum_{j=1,2}\Big\{ -\frac{1}{2}\Delta_{x_j}\otimes \one +\sqrt{\alpha}\lambda_0\int_{\BbbR^3}\frac{\mathrm{d}k}{(2\pi)^{3/2}|k|}\Big[\ex^{\im
 k\cdot x_j}\otimes a(k)+\ex^{-\im k\cdot x_j}\otimes
 a(k)^*\Big]\Big\}\\
&+\frac{\alpha U_0}{|x_1-x_2|}\otimes \one+\one\otimes\Nf.
\end{align*}
$x_j\in\BbbR^3,\, j=1,2,$ are the coordinates of the two electrons. The
electrons are spinless and no statistics is imposed. In addition to the
interaction with the phonons, the electrons repel each other through a
static Coulomb interaction,  which  is proportional to $e^2$. Since
$\sqrt{\alpha}$ is proportional to $e$,   the strength of the
Coulomb repulsion is written as $\alpha U_0$ with $U_0$ a second
dimensionless coupling parameter $U_0\ge 0$. 
As explained  in \cite{DPV}, e.g., 
$U_0\ge \sqrt{2}$ in  the Fr\"ohlich approximation. For the purpose of our study, we regard $\alpha,
U_0$ as independent parameters, $\alpha\ge 0, U_0\ge 0$. 

The phonons induce  an effective attraction between the electrons which
competes with the Coulomb repulsion. If the latter dominates  we expect
the low energy states of $\Hs$ to consist of two far apart polarons,
while  if the coupling to the phonon field dominates the electrons
should form a bound pair.  More precisely. let
$E_{\mathrm{p}}(\alpha)$
and $E_{\mathrm{bp}}(\alpha,U_0)$ be the lowest energy of
$\Hs$ and $\Hb$, respectively. We define the bipolaron  binding energy as 
\begin{align*}
E_{\mathrm{bin}}(\alpha,
 U_0)=2E_{\mathrm{p}}(\alpha)-E_{\mathrm{bp}}(\alpha, U_0).
\end{align*}  
One basic problem is then to characterize in the quadrant of couplings 
$(\alpha, U_0)$ a domain with $E_{\mathrm{bin}}=0$ (two widely separated
polarons) and a domain with $E_{\mathrm{bin}}>0$ (bound pair).

If $\alpha$ is small, one could use iterative  techniques in the
spirit of \cite {BFP}, see also \cite{CHHS, HS}, to approach the issue of a  strictly
positive binding energy. In this paper we investigate  the strong
coupling regime, $\alpha\to \infty$.

We first establish that $\Hs$ is a properly defined self-adjoint
operator and that, for $\Es(\alpha)=\inf\mathrm{spec}(\Hs)$, one has
$\lim_{\alpha\to \infty}\Es(\alpha)/\alpha^2=c_{\mathrm{p}}$ with
$c_{\mathrm{p}}$ a constant defined as the minimum of the Pekar
functional. (Numerically, one finds $c_{\mathrm{p}}=-0.1085$\dots
\cite{Miyake}.)
  The strong coupling limit  has been studied before by Donsker and
Varadhan \cite{DoVa}, using functional integration, and by Lieb and
Thomas \cite{LT1, LT2} based on operator techniques. In fact, we
slightly improve their results. In \cite{DoVa, LT1, LT2} the authors 
consider a suitable cutoff version of $\Hs$ with ground state energy 
$E^{(\kappa)}(\alpha)$, $\kappa$ denoting the ultraviolet cutoff. They
define $E(\alpha)=\lim_{\kappa\to\infty}E^{(\kappa)}(\alpha)$ and prove
that $\lim_{\alpha\to
\infty}E(\alpha)/\alpha^2=c_{\mathrm{p}}$. Secondly
we consider the bipolaron and establish that in the strong coupling
limit its ground state energy is given through minimizing the
Pekar-Tomasevich functional \cite{PT}, see \cite{SF} for a review.
 An analysis of the Pekar-Tomasevich  variational  problem yields
an information on the binding energy for large $\alpha$.

From  our investigaton of the strong coupling limit it is a small step to
study the existence of a ground state for the bipolaron at constant
total momentum $P$ following the strategy developed in \cite{LMS}.
We will prove that, if $E_{\mathrm{bin}}>0$, then $H_{\mathrm{bp}}$ at
total momentum $P$ has a ground state,
provided $P$ is not too large (specified quantitatively).

There is a rich, mostly physics, literature on the bipolaron. We refer
to the listing in \cite{Hirokawa}. Spectral properties of the
Fr\"{o}hlich polaron are investigated in \cite{Moller2, Spohn1}.

The paper is organized as follows: Section 3 deals with the strong limit
$\alpha\to \infty$ and Section 4 with the existence of a ground state.
In the Appendices  A and B removal of the ultraviolet cutoff and
self-adjointness are discussed.

\begin{flushleft}
{\sf Acknowledgements.}
\end{flushleft}
T. Miyao thanks M. Hirokawa for useful comments.

\section{Main Results}
 
In general we denote the inner product and the norm of a Hilbert space
$\h$ by $\la\cdot,\cdot\ra_{\h}$ and $\|\cdot\|_{\h}$ respectively. If
there is no danger of confusion, then we omit the subscript $\h$ in
$\la\cdot,\cdot\ra_{\h}$ and $\|\cdot\|_{\h}$. For a linear operator $T$
on a Hilbert space, we denote its domain by $\D(T)$.
For a self-adjoint operator $A$ on a Hilbert space, we denote its
spectrum (resp. essential spectrum) by $\mathrm{spec}(A)$
(resp. $\mathrm{ess.\, spec}(A)$).

Let $\h$ be a Hilbert space. The Fock space over $\h$ is defined by
\[
 \Fock(\h)=\oplus_{n=0}^{\infty}\otimes_{\mathrm{s}}^n\h,
\]
where $\otimes_{\mathrm{s}}^n\h$ means the $n$-fold symmetric tensor
product of $\h$ with the convention $\otimes_{\mathrm{s}}^0\h=\BbbC$.
The vector $\Omega=1\oplus0\oplus\cdots\in\Fock(\h)$ is called the Fock
vacuum.

We denote by $a(f)$ the annihilation operator on $\Fock(\h)$ with test
vector $f\in\h$ \cite[Sec. X.7]{ReSi2}. By definition, $a(f)$ is densely
defined, closed, and antilinear in $f$. The adjoint $a(f)^*$ is the
adjoint of  $a(f)$ and  called the creation operator. We frequently
write $a(f)^{\#}$ to denote either $a(f)$ or $a(f)^*$.  Creation and
annihilation operators satisfy the canonical
commutation relations
\[
 [a(f),a(g)^*]=\la f,g\ra_{\h}\one,\ \ \ [a(f),a(g)]=0=[a(f)^*,a(g)^*]
\]
on the finite particle subspace
\[
 \Fock_0(\h)=\bigcup_{m=1}^\infty\{\varphi=\varphi_0\oplus\varphi_1\oplus\cdots
 \in\Fock(\h)\, |\, \varphi_n=0,\ \mathrm{for}\, n\ge m \},
\]
where 
$\one$ denotes the identity operator.
In the case of $\h=L^2(\BbbR^3)$, we often use the symbolic notation for
the annihilation and
creation operator by the kernel:
\[
 a(f)=\int_{\BbbR^3}\mathrm{d}k\, f(k)^*a(k),\ \
 a(f)^*=\int_{\BbbR^3}\mathrm{d}k\, f(k)a(k)^*.
\] 
We  introduce a further  important
subspace of $\Fock(\h)$. Let  $\mathfrak{s}$ be a subspace of $\h$.
We define
\[
 \Ffin(\mathfrak{s})=\mathrm{Lin}\{a(f_1)^*\dots a(f_n)^*\Omega,\
 \Omega\, |\, f_1,\dots,f_n\in\mathfrak{s},\ n\in\BbbN\},
\]
where $\mathrm{Lin}\{\cdots\}$ means the linear span  of the  set
$\{\cdots\}$.
If $\mathfrak{s}$ is dense in $\h$, so is $\Ffin(\mathfrak{s})$ in $\Fock(\h)$.

Let $b$ be a contraction operator from $\h_1$ to $\h_2$, i.e., $\|b\|\le 1$.
The linear operator $\Gamma(b):\Fock(\h_1)\to \Fock(\h_2)$ is defined by
\[
 \Gamma(b)\upharpoonright\otimes_\mathrm{s}^n\h_1=\otimes^n b
\]
with the convention $\otimes^0 b=\one$.

For a densely defined closable operator $c$ on $\h$, $\dG(c):\Fock(\h)\to
\Fock(\h)$ is defined by
\[
 \dG(c)\upharpoonright \hotimes^n_{\mathrm{s}}\D(c)=\sum_{j=1}^{n}
\one\otimes\cdots\otimes\underset{j\,  \mathrm{th}}{c}\otimes \cdots\otimes\one
\]
and 
\[
 \dG(c)\Omega=0
\]
 where $\hotimes$ 
means the algebraic tensor product.
Here in the $j$-th summand $c$ is at the $j$-th entry.
Clearly $\dG(c)$ is closable and we denote its closure by the same
symbol. As a typical   example,  the number operator $N_{\mathrm{f}}$ is given
by  $N_{\mathrm{f}}=\dG(\one)$.

The bipolaron Hamiltonian with an ultraviolet cutoff $\kappa>0$ is
defined as 
\begin{align*}
&\Hbk\\
&=\sum_{j=1,2}\Big\{ -\frac{1}{2}\Delta_{x_j}\otimes \one +\sqrt{\alpha}\lambda_0\int_{|k|\le
 \kappa}\frac{\mathrm{d}k}{(2\pi)^{3/2}|k|}\Big[\ex^{\im
 k\cdot x_j}\otimes a(k)+\ex^{-\im k\cdot x_j}\otimes
 a(k)^*\Big]\Big\}\\
&\ \ \ +\frac{\alpha U_0}{|x_1-x_2|}\otimes \one+\one\otimes\Nf
\end{align*}
with $\alpha, U_0\ge 0$.
 This linear operator  acts in the Hilbert
space $L^2(\BbbR^6, \mathrm{d}x_1\otimes\mathrm{d}x_2)\otimes \Fock(L^2(\BbbR^3))$.
By the bound
\begin{align}
\|a(f)^{\#}(\Nf+\one)^{-1/2}\|\le \|f\|\label{NumberEst}
\end{align}
 and the Kato-Rellich theorem,
it is easy to see that, for all $0<\kappa<\infty$ and $0<\alpha<\infty$,  $\Hbk$ is
self-adjoint on the domain of the  self-adoint operator
$L_{\mathrm{bp}}=-\sum_{j=1,2}\Delta_{x_j}\otimes\one+\one\otimes \Nf$, bounded from
below, and essentially self-adjoint on any core for   $L_{\mathrm{bp}}$. 
We note  that $\Hbk$ strongly commutes with the total
momentum operator
\begin{align}
 \Pt=-\im\nabla_{x_1}\otimes\one-\im\nabla_{x_2}\otimes\one+\one\otimes
 P_{\mathrm{f}},
\end{align}
where $P_{\mathrm{f}}=(\dG(k_1),\dG(k_2),\dG(k_3))$, that is to say,
$\ex^{\im a\cdot \Pt}\Hbk\subseteq \Hbk\ex^{\im a\cdot \Pt}$ for all $a\in\BbbR^3$.

Let $(\xr, \xc)$ be the center of mass coordinates defined by
\[
 \xr=x_1-x_2,\ \ \ \xc=\frac{x_1+x_2}{2}
\]
and let $U_C$ be the unitary operator from $L^2(\BbbR^6,
\mathrm{d}x_1\otimes \mathrm{d}x_2)$ to $L^2(\BbbR^6,
\mathrm{d}\xr\otimes\mathrm{d}\xc)$ given by
\[
 (U_Cf)(\xr,\xc)=f\big(\xc+\frac{\xr}{2}, \xc-\frac{\xr}{2}\big)
\]
for $f(x_1,x_2)\in L^2(\BbbR^6,\mathrm{d}x_1\otimes\mathrm{d}x_2)$.
We introduce a unitary operator $U$ by
\[
 \uni=(\mathcal{F}_{\xc}\otimes\one)\,  \ex^{\im \xc\cdot P_{\mathrm{f}}}(U_C\otimes\one),
\]
where $\mathcal{F}_{\xc}$ is the Fourier transformation with respect  to
$\xc$, i.e.,
\[
 (\mathcal{F}_{\xc}f)(P,\xr)=(2\pi)^{-3/2}\int _{\BbbR^3}\mathrm{d}\xc\, \ex^{-\im \xc\cdot
 P}f(\xr, \xc)
\]
for $f(\xr,\xc)\in L^2(\BbbR^6, \mathrm{d}\xr\otimes\mathrm{d}\xc)$.
The unitary operator $\uni$ induces the identification $\FullHil$ with
$\int^{\oplus}_{\BbbR^3}\OneHil\, \mathrm{d}P$,  that is
concretely written as 
\begin{align*}
 &(\uni\vphi)^{(n)}(P,\xr,k_1,\dots,k_n)\\
=&(2\pi)^{-3/2}\int_{\BbbR^3}\mathrm{d}\xc\, \ex^{-\im
 \xc\cdot(P-\sum_{j=1}^nk_j)}\vphi^{(n)}\Big(\xc+\frac{\xr}{2},\xc-\frac{\xr}{2},k_1,\dots,k_n\Big)
\end{align*}
for $\vphi=\oplus_{n=0}^{\infty}\vphi^{(n)}\in \FullHil$.
It is easily shown that 
\[
 \uni \Pt \uni^*=\int^{\oplus}_{\BbbR^3}P\, \mathrm{d}P.
\] 
Hence the unitary operator $\uni$ provides the direct integral
decomposition of\\ $\FullHil$ with respect to the value of the total momentum. 

Since $\Hbk$ strongly commutes with $\Pt$, $\uni\Hbk \uni^*$ is decomposable
and can be represented by the fiber direct integral 
\[
 \uni\Hbk \uni^*=\int^{\oplus}_{\BbbR^3}H_{\kappa}(P)\, \mathrm{d}P,
\]
where
\begin{align}
H_{\kappa}(P)=&\frac{1}{4}(P-\one\otimes
 \Pf)^2-\Delta_{\xr}\otimes\one+\frac{\alpha U_0}{|\xr|}\otimes\one+\one\otimes\Nf\nonumber\\
&+2\sqrt{\alpha}\lambda_0\int_{|k|\le \kappa}\frac{\mathrm{d}k}{(2\pi)^{3/2}|k|}\cos\frac{k\cdot
 \xr}{2}\otimes\big[a(k)+a(k)^*\big].
\end{align}
 By the Kato-Rellich's theorem,
$H_{\kappa}(P)$ is self-adjoint on
$\D(-\Delta_{\xr}\otimes\one)\cap\D(\one\otimes\Pf^2)\cap\D(\one\otimes\Nf)$
for all $\kappa<\infty$ and $\alpha<\infty$, and bounded from
below. Further, $H_{\kappa}(P)$ is essentially self-adjoint on any core
for the self-adjoint operator 
\begin{align}
 L=-\Delta_{\xr}\otimes\one+\one\otimes\Pf^2+\one\otimes\Nf.\label{LOp}
\end{align}

We  state our main results. Our first result concerns
the existence of the limiting Hamiltonians. Namely, we  remove the
ultraviolet cutoff  from $\Hbk$ and $H_{\kappa}(P)$ without  energy renormalization.

\begin{Thm}\label{DefHami}
\begin{itemize}
\item[{\rm (i)}] For all $\alpha<\infty$ and $U_0<\infty$, there exists a self-adjoint
		 operator $\Hb$ that is bounded from below such that 
$\Hbk$ converges to $\Hb$ in the strong resolvent sense.
\item[{\rm (ii)}] For all $\alpha<\infty$, $U_0<\infty$ and $P\in\BbbR^3$, there
		 exists a self-adjoint
operator $H(P)$ that is bounded from below such that $H_{\kappa}(P)$
		 converges to $H(P)$ in the strong resolvent sense.
\item[{\rm (iii)}] $\uni\Hb \uni^*$ is decomposable and 
\begin{align}
\uni\Hb \uni^*=\int^{\oplus}_{\BbbR^3}H(P)\, \mathrm{d}P.
\end{align}  
\end{itemize}
\end{Thm}

Let $\Hsk$ be the Hamiltonian for a  single  polaron  with the  ultraviolet
cutoff $\kappa$,
\begin{align*}
&\Hsk\\
&=-\frac{1}{2}\Delta_x\otimes\one+\sqrt{\alpha}\lambda_0
 \int_{|k|\le \kappa}\frac{\mathrm{d}k}{(2\pi)^{3/2}|k|}
\Big[\ex^{\im k\cdot x}\otimes a(k)+\ex^{-\im k\cdot x}\otimes
 a(k)^*\Big]+\one\otimes \Nf.
\end{align*}
 The linear operator $\Hsk$ 
acts in the Hilbert space $L^2(\BbbR^3)\otimes
\Fock(L^2(\BbbR^3))$. Moreover, for all $0<\kappa<\infty$ and
$0<\alpha<\infty$, $\Hsk$ is self-adjoint on the domain of the
self-adjoint operator $L_{\mathrm{p}}=-\Delta_x\otimes\one +\one \otimes
\Nf$, bounded from below, and essentially self-adjoint on any core for
$L_{\mathrm{p}}$. In  a  way  similar to  the proof of  Theorem
\ref{DefHami} (i), we can  show the following.

\begin{Prop}
For any coupling $\alpha$, there exists a self-adjoint operator $\Hs$, bounded from below,  such that $\Hsk$ converges to
 $\Hs$ in the strong  resolvent sense as $\kappa\to \infty$. 
\end{Prop}

 Let
\[
 E_{\mathrm{bp}}=\inf \mathrm{spec}(\Hb),\ \
 E_{\mathrm{p}}=\inf \mathrm{spec}(\Hs).
\]
The {\it binding energy} $E_{\mathrm{bin}}$ is defined by
\[
 E_{\mathrm{bin}}=2E_{\mathrm{p}}-E_{\mathrm{bp}}.
\]
In order to display  the dependence on  $\alpha$ and $U_0$, we also  denote the binding
energy by $E_{\mathrm{bin}}(\alpha, U_0)$.

We introduce the {\it Pekar energy functional} by
\begin{align}
\ep(\vphi)=\frac{1}{2}\int \mathrm{d}x\, |\nabla_x
 \vphi(x)|^2-\frac{1}{\sqrt{2}}\int\mathrm{d}x\mathrm{d}y\,
 \frac{|\vphi(x)|^2|\vphi(y)|^2}{|x-y|}
\end{align}
for $\vphi\in W^1(\BbbR^3)$, where  $W^1(\BbbR^d)$  is the space of functions on
 $\BbbR^d$  such  that
 $\|\nabla \vphi\|_{L^2(\BbbR^d)}$ and $\|\vphi\|_{L^2(\BbbR^d)}$
 are finite.
For $U\ge 0$, the {\it Pekar-Tomasevich energy  functional} is defined by
\begin{align}
\eb^U(\vphi)
=&\frac{1}{2}\int \mathrm{d}x_1\mathrm{d}x_2\, |\nabla_{x_1}
 \vphi(x_1,x_2)|^2+\frac{1}{2}\int \mathrm{d}x_1\mathrm{d}x_2\, |\nabla_{x_2}
 \vphi(x_1,x_2)|^2\nonumber\\
&+U\int\mathrm{d}x_1\mathrm{d}x_1\, \frac{|\vphi(x_1,
 x_2)|^2}{|x_1-x_2|}\nonumber\\
&-\frac{1}{\sqrt{2}}\sum_{i,j=1,2}\int\mathrm{d}x_1\mathrm{d}x_2\mathrm{d}y_1\mathrm{d}y_2\,
 \frac{|\vphi(x_1,x_2)|^2|\vphi(y_1,y_2)|^2}{|x_i-y_j|}
\end{align} 
for $\vphi\in W^1(\BbbR^6)$.

\begin{Thm}\label{Binding}
Let 
\begin{align}
c_{\mathrm{p}}&=\inf\{\ep(\vphi)\, |\, \vphi\in W^1(\BbbR^3),\,
 \|\vphi\|_{L^2(\BbbR^3)}=1\},\\
c_{\mathrm{bp}}(U)&=\inf\{\eb^U(\vphi)\, |\, \vphi\in W^1(\BbbR^6),\, \|\vphi\|_{L^2(\BbbR^6)}=1\}.
\end{align}
 For any Coulomb strength $U_0\ge 0$,
\[
\lim_{\alpha\to\infty}\frac{E_{\mathrm{bin}}(\alpha, U_0)}{
 \alpha^2}=2c_{\mathrm{p}}-c_{\mathrm{bp}}
(U_0).
\]
\end{Thm}
The Pekar energy functional is studied in \cite{Lieb}. In a separate work
\cite{MS} we investigate the Pekar-Tomasevich energy  functional and quote only

\begin{Thm}{\rm \cite{MS}}\label{PropertPekar}
\begin{itemize}
\item[{\rm (i)}]For all $U\ge 0$,
		$2c_{\mathrm{p}}-c_{\mathrm{bp}}(U)\ge 0$. Moreover,
		$2c_{\mathrm{p}}-c_{\mathrm{bp}}(U)$ is monotone
		decreasing, convex  and continuous in $U$.
\item[{\rm (ii)}] Let $U_{\mathrm{c}}=\sup\{U\in [0, \infty)\, |\, 
		2c_{\mathrm{p}}-c_{\mathrm{bp}}(U)>0\}$. Then $
						     \sqrt{2}< U_{\mathrm{c}}.
						    $
\end{itemize}
\end{Thm}
\begin{rem}
{\rm
 If $\vphi(x_1,x_2)=\phi_0(x_1)\phi_0(x_2)$ with $\phi_0$ the
	   minimizer
of $\ep(\cdot)$, up to translation, then
 $\eb^{\sqrt{2}}(\vphi)=2c_{\mathrm{p}}$. 
Theorem \ref{PropertPekar}  asserts  that  the energy is lowered through correlations.
Numerically one uses trial functions \cite{SD} or variational actions
 \cite{DPV}. On this basis the accepted value for $U_{\mathrm{c}}$ is
 approximately $(1.1)\sqrt{2}$.
}
\end{rem}

Returning to finite $\alpha$ we  characterize the existence of the ground state for $H(P)$ in
terms of the binding energy in the  following way.

\begin{Thm}\label{GapSpec}
For all $P$, coupling strength $\alpha$ and  Coulomb strength $U_0$, one
 has 
\begin{align*}
\inf \mathrm{ess.\, spec}(H(P))-\inf\mathrm{spec}(H(P))
\ge \min\big\{1,E_{\mathrm{bin}}\big\}-\frac{P^2}{4}.
\end{align*}
Thus, if  $E_{\mathrm{bin}}>0$,  then $H(P)$ has a ground state provided
\[
 |P|<2\min\big\{1, \sqrt{E_{\mathrm{bin}}}\big\},
\]
. 
\end{Thm}

Combining both theorems yields a domain of coupling parameters  and $P$ for
which $H(P)$ has a ground state.
\begin{coro}\label{ExistenceGS}Suppose that the strength $U_0$ of the Coulomb
 interaction  satisfies $U_0< U_{\mathrm{c}}$.
 Then,
there exists an  $\alpha_{\mathrm{c}}$ such that, for any $\alpha> \alpha_{\mathrm{c}}$,
 $H(P)$  has a ground state for $|P|<2$.
\end{coro}

\section{Strong coupling limit}
\subsection{The Pekar variational problem}
In this seubsection we summarize properties of  the
Pekar-Tomasevich energy  functional.  They  are proven in \cite{MS}.

\begin{lemm}
\begin{itemize}
\item[{\rm (i)}] For all $\vphi\in W^1(\BbbR^3)$ with $\|\vphi\|_{L^2(\BbbR^3)}=1$,
		 there exists a constant $A_{\mathrm{p}}>-\infty$ such
		 that $\ep(\vphi)\ge A_{\mathrm{p}}$. Hence,
		 $c_{\mathrm{p}}>-\infty$.
\item[{\rm (ii)}] For all $\vphi\in W^1(\BbbR^6)$ with
		 $\|\vphi\|_{L^2(\BbbR^6)}=1$, there exists a constant
		 $A_{\mathrm{bp}}>-\infty$ such that $\eb^U(\vphi)\ge
		 A_{\mathrm{bp}}$. Hence, $c_{\mathrm{bp}}(U)>-\infty$.
\end{itemize}
\end{lemm}

\begin{lemm}\label{CApp}
\begin{itemize}
\item[{\rm (i)}] $\displaystyle c_{\mathrm{p}}=\inf\big\{\ep(\vphi)\, |\, \vphi\in
		 C_0^{\infty}(\BbbR^3),\,
		 \|\vphi\|_{L^2(\BbbR^3)}=1\big\}$.
\item[{\rm (ii)}] $\displaystyle c_{\mathrm{bp}}(U)=\inf\big\{\eb^U(\vphi)\, |\, \vphi\in
		 C_0^{\infty}(\BbbR^6),\,
		 \|\vphi\|_{L^2(\BbbR^6)}=1\big\}$ for all $U\ge 0$.
\end{itemize}
\end{lemm}

\begin{lemm}\label{CContinuous}
$c_{\mathrm{bp}}(U)$ is continuous in $U\ge 0$.
\end{lemm}

\subsection{Infinimum of spectrum for $\alpha\to \infty$}

\begin{lemm}\label{Upperbound}For all $\alpha>0$ and Coulomb strength $U_0$, we have the following.
\begin{itemize}
\item[\rm{(i)}] $\Es\le c_{\mathrm{p}}\alpha^2$.
\item[\rm{(ii)}] $\Eb\le c_{\mathrm{bp}}(U_0)\alpha^2$.
\end{itemize}
\end{lemm}
{\it Proof.} (i) We will apply the variational principle. Let $\vphi\in
C_0^{\infty}(\BbbR^3)$ with $\|\vphi\|_{L^1(\BbbR^3)}=1$. Set 
\[
 \rho(k)=\frac{1}{(2\pi)^{3/2}}\int \mathrm{d}x\, \ex^{-\im k\cdot x}|\vphi(x)|^2.
\]
We choose $\xi=\vphi\otimes \Psi$ as a trial function, where
\begin{align*}
\Psi=\exp\Big\{\im\lambda\int_{|k|\le
 \kappa}\frac{\mathrm{d}k}{|k|}\Big[-\im \bar{\rho}(k)a(k)+\im\rho(k)a(k)^*\Big]\Big\}\Omega
\end{align*}
with $\lambda=\sqrt{\alpha}\lambda_0$.
By the standard calculation, we have
\[
 \la \xi, \Hsk\xi\ra=\frac{1}{2}\int\mathrm{d}x\, |\nabla_x
 \vphi(x)|^2-\lambda^2\int_{|k|\le \kappa}\mathrm{d}k\, \frac{|\rho(k)|^2}{|k|^2}
.\]
Thus 
\[
 \Esk\le \frac{1}{2}\int\mathrm{d}x\, |\nabla_x
 \vphi(x)|^2-\lambda^2\int_{|k|\le \kappa}\mathrm{d}k\, \frac{|\rho(k)|^2}{|k|^2}
.
\]
Here  $\Esk$ is  the ground state energy for  $\Hsk$.
Taking the limit $\kappa\to \infty$, we have 
\begin{align*}
 \Es&\le \frac{1}{2}\int\mathrm{d}x\, |\nabla_x
 \vphi(x)|^2-\lambda^2\int\mathrm{d}k\, \frac{|\rho(k)|^2}{|k|^2}\\
&=\frac{1}{2}\int\mathrm{d}x\, |\nabla_x\vphi(x)|^2-\frac{\alpha}{\sqrt{2}}\int\mathrm{d}x\mathrm{d}y\frac{|\vphi(x)|^2|\vphi(y)|^2}{|x-y|}
\end{align*}
by Proposition \ref{LimitEnergy} (ii).
Here  we use the following fact:
\begin{align}
\int_{\BbbR^3}\mathrm{d}k\frac{\bar{\hat{f}}(k)\hat{g}(k)}{k^2}=\frac{1}{4\pi}\int_{\BbbR^3}\int_{\BbbR^3}\mathrm{d}x\mathrm{d}y\, \frac{\bar{f}(x)g(y)}{|x-y|},\label{Fourier}
\end{align}
for $f,g\in L^{6/5}(\BbbR^3)$, where
$\hat{f}(k)=(2\pi)^{-3/2}\int_{\BbbR^3}\mathrm{d}x\, \ex^{-\im k\cdot x}f(x)$.
 Finally we remark that, by the scaling argment and Lemma \ref{CApp} (i), we
get
\begin{align*}
 &\inf\Big\{\frac{1}{2}\int\mathrm{d}x\, |\nabla_x
 \vphi(x)|^2-\frac{\alpha}{\sqrt{2}}\int\mathrm{d}x\mathrm{d}y\,
 \frac{|\vphi(x)|^2|\vphi(y)|^2}{|x-y|}\, \Big|\, \vphi\in
 C^{\infty}_0(\BbbR^3),\, \|\vphi\|_{L^2}=1\Big\}\\
&=c_{\mathrm{p}}\alpha^2.
\end{align*}

(ii) The proof of (ii) is almost same as (i). Our choice of the trial
function is 
\begin{align*}
\xi=&\vphi\otimes \Psi,\ \ \ \vphi\in C_0^{\infty}(\BbbR^6)\ \
 \mbox{with}\ \  \|\vphi\|_{L^2(\BbbR^6)}=1,\\
\Psi=&\exp\Big\{\im\lambda\int_{|k|\le
 \kappa}\frac{\mathrm{d}k}{|k|}\Big[-\im
 \bar{\rho}(k)a(k)+\im\rho(k)a(k)^*\Big]\Big\}\Omega\ \ \mbox{with}\ \ \lambda=\sqrt{\alpha}\lambda_0,\\
\rho(k)=&\rho_1(k)+\rho_2(k),\\
\rho_1(k)=&\frac{1}{(2\pi)^{3/2}}\int \mathrm{d}x_1\mathrm{d}x_2\,
 \ex^{-\im k\cdot x_2}|\vphi(x_1, x_2)|^2,\\
\rho_2(k)=&\frac{1}{(2\pi)^{3/2}}\int \mathrm{d}x_1\mathrm{d}x_2\,
 \ex^{-\im k\cdot x_1}|\vphi(x_1, x_2)|^2.
\end{align*}
Then, we get 
\[
 \la \xi,
 \Hbk\xi\ra=T_{\mathrm{bp}}(\vphi)+\alpha U_0\int\mathrm{d}x_1\mathrm{d}x_2\,
 \frac{|\vphi(x_1,x_2)|^2}{|x_1-x_2|}-\lambda^2\int_{|k|\le
 \kappa}\mathrm{d}k\, \frac{|\rho(k)|^2}{k^2}.
\]
Accordingly, by Proposition \ref{LimitEnergy} (i), we obtain that 
\[
 \Eb\le T_{\mathrm{bp}}(\vphi)+\alpha U_0 \int\mathrm{d}x_1\mathrm{d}x_2\,
 \frac{|\vphi(x_1,x_2)|^2}{|x_1-x_2|}-\lambda^2\int\mathrm{d}k\, \frac{|\rho(k)|^2}{k^2}.
\]
Let $\rho_1(k; x_1):=(2\pi)^{-3/2}\int\mathrm{d}x_2\, \ex^{-\im k\cdot
x_2}|\vphi(x_1,x_2)|^2$.
Then, by Fubini's theorem and (\ref{Fourier}),
\begin{align*}
\lambda^2\int\mathrm{d}k\,
 \frac{|\rho_1(k)|^2}{k^2}=&\lambda^2\int\mathrm{d}x_1\mathrm{d}y_1\int\mathrm{d}k\,
 \frac{\bar{\rho}_1(k;x_1)\rho_1(k;y_1)}{k^2}\\
=&\lambda^2\int\mathrm{d}x_1\mathrm{d}y_1\Big(\frac{1}{4\pi}\int\mathrm{d}x_2\mathrm{d}y_2\frac{|\vphi(x_1,x_2)|^2|\vphi(y_1,y_2)|^2}{|x_2-y_2|}\Big)\\
=&-\alpha W^{(2,2)}_{\mathrm{bp}}(\vphi).
\end{align*}
Calculating the other terms  contained in  $\lambda^2\int\mathrm{d}k\, |\rho(k)|^2/k^2$
by the similar way, we obtain 
\[
 -\lambda^2\int\mathrm{d}k\frac{|\rho(k)|^2}{k^2}=\alpha W_{\mathrm{bp}}(\vphi).
\]
Now the assertion follows from Lemma \ref{CApp} (ii) and the scaling
argument. $\Box$

\begin{lemm}\label{Lowerbound}
\begin{itemize}
\item[{\rm (i)}]$\Es\ge c_{\mathrm{p}}\alpha^2+\mathcal{O}(\alpha^{9/5})$.
\item[{\rm (ii)}]$\Eb\ge
		c_{\mathrm{bp}}\Big((1-c_1\alpha^{-1/5})(1-c_2\alpha^{-1/5})U_0\Big)\alpha^2+\mathcal{O}(\alpha^{9/5})$,
		where $c_1$ and $c_2$ are positive constants.
\end{itemize}
\end{lemm}
{\it Proof.} The assertion (i) has been  proven in \cite{LT1,
LT2}, essentially. Although the authors consider a  finite volume model,
 their arguments are   still valid  in our case. More precisely, first we
apply the methods in \cite{LT1, LT2} to $\Hsk$ for sufficiently large
$\kappa$, and  obtains that
\[
 \Esk\ge c_{\mathrm{p}}\alpha^2+\mathcal{O}(\alpha^{9/5}).
\]
The important point is that the error term $\mathcal{O}(\alpha^{9/5})$ is independent of
$\kappa$. Now taking the limit $\kappa\to \infty$, we have the desired
result by Proposition \ref{LimitEnergy} (ii).
As for (ii), one  can extend the proof of (i) to the bipolaron Hamiltonian
$\Hbk$ with some slight modifications. $\Box$
\medskip\\\medskip\\
{\large{\sf Proof of Theorem \ref{Binding}}}\\
By Lemma \ref{Upperbound} and \ref{Lowerbound}, we have
\begin{align*}
&2c_{\mathrm{p}}\alpha^2-c_{\mathrm{bp}}\Big((1-c_1\alpha^{-1/5})(1-c_2\alpha^{-1/5})U_0\Big)\alpha^2+\mathcal{O}(\alpha^{9/5})\\
\ge& 2\Es-\Eb\\
\ge& 2c_{\mathrm{p}}\alpha^2-c_{\mathrm{bp}}(U_0)\alpha^2+\mathcal{O}(\alpha^{9/5}).
\end{align*}
Taking Lemma \ref{CContinuous}  into consideration,
we get
\[
 \lim_{\alpha\to\infty}\frac{E_{\mathrm{bin}}(\alpha, U_0)}{\alpha^2}=2c_{\mathrm{p}}-c_{\mathrm{bp}}(U_0).
\ \ \ \Box
\]

\section{Existence of a ground state}
\subsection{Properties of the ground state energy}
Let $\Ebk$ and $\Esk$ be the ground state energy for $\Hbk$ and $\Hsk$
 respectively.
Further we denote $\inf \mathrm{spec}(H_{\kappa}(P))$,
resp. $\inf\mathrm{spec}(H(P))$,  by $E_{\kappa}(P)$, resp. $E(P)$.

\begin{Prop}\label{PropEnergy}
For all $\alpha, U_0>0$ and $\kappa\le \infty$, the following holds.
\begin{itemize}
\item[{\rm (i)}]
$\displaystyle
E_{\kappa}(P)\le E_{\kappa}(0)+\frac{P^2}{4}
$ for all $P$.
\item[{\rm (ii)}] $\displaystyle E_{\kappa}(0)\le E_{\kappa}(P)$ for all
		$P$.
\item[{\rm (iii)}] $\displaystyle E_{\kappa}(0)=\Ebk$.
\end{itemize}
\end{Prop}
{\it Proof.} These are well-known relations. However, for the  reader's
convenience, we give a proof.

(i) Let $T$ be the time reversal operator which is defined by complex
conjugation  the wave function, reversing all phonon momenta. $T$ is
antiunitary and $TH_{\kappa}(P)T=H_{\kappa}(-P)$. Thus we conclude that 
\begin{align}
E_{\kappa}(-P)=E_{\kappa}(P).\label{SymmEnergy}
\end{align}

Let $F(P):=E_{\kappa}(P)-P^2/4$. Then,
 it is clear that $F$ is concave. Moreover, by (\ref{SymmEnergy}), 
$F(-P)=F(P)$. Thus,
\[
 F(0)=F\big(\frac{P}{2}-\frac{P}{2}\big)\ge \frac{1}{2}F(P)+\frac{1}{2}F(-P)=F(P).
\]

(ii) Let 
\[
 K(P)=\frac{1}{4}(P-\one\otimes\Pf)^2
\]
and 
\begin{align*}
 H=&-\Delta_{\xr}\otimes\one+\frac{\alpha U_0}{|\xr|}\otimes\one+\one\otimes\Nf\\
&+2\sqrt{\alpha}\lambda_0\int_{|k|\le
 \kappa}\frac{\mathrm{d}k}{(2\pi)^{3/2}|k|}\cos\frac{k\cdot\xr}{2}\otimes[a(k)+a(k)^*].
\end{align*}
Then, $H_{\kappa}(P)=K(P)\dot{+}H$, where $\dot{+}$ means the form sum.
We consider the Schr\"{o}dinger representation $L^2(Q, \mathrm{d}\mu)$
of the Fock space $\Fock(L^2(\BbbR^3))$, where $\mathrm{d}\mu$ is the
Gaussian measure with mean $0$ and covariance $1/2$. Let
$\vth$ be the unitary operator which gives the natural identification
from $L^2(\BbbR^3)\otimes\Fock(L^2(\BbbR^3))$ onto $L^2(\BbbR^3\times Q,\mathrm{d}\xr\otimes \mathrm{d}\mu)$. We note that
$\vth\ex^{-tH}\vth^*$ is positivity preserving, see e.g., \cite{BFS}. 
Moreover, since 
\[
 \ex^{-t(P_j-\one\otimes P_{\mathrm{f},j})^2/4}=\int\mathrm{d}\mu_{G}
 (\lambda)\, \ex^{\im
 \lambda(P_j-\one\otimes P_{\mathrm{f},j})},\ \ \ j=1,2,3,
\]
where $\mu_G$ is the Gaussian measure with mean zero and variance $t/2$,
and $\vth \ex^{-\im\lambda P_{\mathrm{f},j}}\vth^*$ is positivity
preserving (see, e.g., \cite{Simon}), we get 
\begin{align*}
 |\vth \ex^{-t(P_j-\one\otimes P_{\mathrm{f},j})^2/4}\vth^*\vphi|
\le &\int\mathrm{d}\mu_G(\lambda)\, |\vth\ex^{-\im\lambda\one\otimes
 P_{\mathrm{f},j}}\vth^*\vphi|\\
 \le&\int\mathrm{d}\mu_G(\lambda)\, \vth\ex^{-\im\lambda\one\otimes
 P_{\mathrm{f},j}}\vth^*|\vphi|\\
\le&
  \vth \ex^{-t \one\otimes P_{\mathrm{f},j}}\vth^*|\vphi|.
\end{align*}
(Here we use the following fact: if $A$ is positivity preserving, then
$|A\vphi|\le A|\vphi|$.)
Therefore we conclude that 
\begin{align}
|\vth \ex^{-tK(P)}\vth^*\vphi|\le \vth\ex^{-tK(0)}\vth^*|\vphi|.\label{KInq}
\end{align}
Let $T_n=(\ex^{-tK(P)/n}\ex^{-tH/n})^n$. By the Trotter product formula,
$\slim T_n(P)=\ex^{-tH_{\kappa}(P)}$. On the other hand, by the positivity
preserving property for $\vth\ex^{-tH}\vth^*$ and  (\ref{KInq}), we get
$|\vth T_n(P)\vth^*\vphi|\le \vth T_n(0)\vth^*|\vphi|$. Taking the
limit $n\to \infty$, we arrive at $|\vth \ex^{-t
H_{\kappa}(P)}\vth^*\vphi|\le \ex^{-tH_{\kappa}(0)}|\vphi|$ which
implies that 
\begin{align}
 \la \vphi, \vth\ex^{-tH_{\kappa}(P)}\vth^*\vphi\ra\le \la|\vphi|, \vth\ex^{-tH_{\kappa}(0)}\vth^*|\vphi|\ra. \label{EnergyInq}
\end{align}
Now we can easily derive (iii) from the above inequality.

(iii) To show $E_{\kappa}(0)\ge \Ebk$ is easy.  To prove the converse, we
 just note that, by (ii), 
\begin{align*}
\la \vphi, \Hbk\vphi\ra=&\int\mathrm{d}P\, \big\la (\uni\vphi)(P),
 H_{\kappa}(P)(\uni\vphi)(P)\big\ra_{L^2(\BbbR^3)\otimes\Fock(L^2(\BbbR^3))}\\
\ge &\int\mathrm{d}P
 E_{\kappa}(P)\big\|(\uni\vphi)(P)\big\|^2_{L^2(\BbbR^3)\otimes\Fock(L^2(\BbbR^3))}\\
\ge& E_{\kappa}(0)\|\vphi\|^2.\ \ \ \Box
\end{align*}

\subsection{Properties of the ionization  energy}

We introduce the ionization  energy $\Sigma_{\kappa}(P)$ by
\begin{align*}
\Sigma_{\kappa}(P)=\lim_{R\to\infty}\inf_{\vphi\in\mathcal{D}_R,\, 
 \|\vphi\|=1}\la\vphi, H_{\kappa}(P)\vphi\ra,
\end{align*}
where $\mathcal{D}_R=\{\vphi\in\D(H_{\kappa}(P))\, |\,
\vphi(x)=0\ \mbox{if $|\xr|< R$}\}$.

\begin{Prop}\label{PropThreshold}
For all  $\alpha, U_0>0$ and $\kappa<\infty$, the following holds.
\begin{itemize}
\item[{\rm (i)}] $\displaystyle \Sigma_{\kappa}(P)\ge
		 \Sigma_{\kappa}(0)$ for all $P$.
\item[{\rm (ii)}] $\displaystyle \Sigma_{\kappa}(0)=2\Esk$.
\end{itemize}
\end{Prop}
{\it Proof.} (i) We consider the Schr\"{o}dinger representation introduced
 in the previous subsection.
 By (\ref{EnergyInq}), we have
\begin{align*}
\frac{1}{t}\Big\la\vphi, \Big(\one-\vth
 \ex^{-t(H_{\kappa}(P)-E_{\kappa}(0))}\vth^*\Big)\vphi\Big\ra
\ge \frac{1}{t}\Big\la|\vphi|, \Big(\one-\vth
 \ex^{-t(H_{\kappa}(0)-E_{\kappa}(0))}\vth^*\Big)|\vphi|\Big\ra\ge 0
\end{align*}
for all $t>0$.  By taking the limit $t\searrow 0$, we can conclude that
if $\vphi\in \\ \D(\vth |H_{\kappa}(P)|^{1/2}\vth^*)$, then $|\vphi|\in \D(\vth
|H_{\kappa}(0)|^{1/2}\vth^*)$ and 
\begin{align}
\big\la \vphi, \vth H_{\kappa}(P)\vth^*\vphi\big\ra \ge \big\la |\vphi|, \vth H_{\kappa}(0)\vth^*|\vphi|\big\ra\label{ThresholdInq}
\end{align}
as an inequality of forms. Let
\begin{align}
& \tilde{\Sigma}_{R,\kappa}(P)\nonumber\\
=&\inf\Big\{\la \vphi,
 H_{\kappa}(P)\vphi\ra\, \big|\, \vphi\in\D(|H_{\kappa}(P)|^{1/2}),\,
 \|\vphi\|=1\ \mbox{and}\ \vphi(\xr)=0\ \mbox{if}\ |\xr|<R\Big\}.\label{modifiedThresh}
\end{align} 
Then, by (\ref{ThresholdInq}), we get 
\begin{align}
\tilde{\Sigma}_{R,\kappa}(P)\ge
\tilde{\Sigma}_{R,\kappa}(0).\label{FormIonization}
\end{align}
Since, by Lemma \ref{EqSigma} below,
$\lim_{R\to\infty}\tilde{\Sigma}_{R,\kappa}(P)=\Sigma_{\kappa}(P)$,
we get the desired assertion. 

(ii) Let
\begin{align*}
\Sigma(\Hbk)
=&\lim_{R\to\infty}\inf\Big\{\la \vphi, \Hbk\vphi\ra\, \big|\,
 \vphi\in\D(\Hbk), \|\vphi\|=1\\
&\hspace{1cm}\ \mbox{and}\ \vphi(x_1,x_2)=0\ \mbox{if}\ |x_1-x_2|<R\Big\}.
\end{align*}
The inequality $\Sigma_{\kappa}(P)\ge \Sigma(\Hbk)$ has been essentially
proven in \cite{FGS2}. Namely assume that there exists $P_0$ such that
$\Sigma_{\kappa}(P_0)<\Sigma(\Hbk)$. Then there exists an $R>0$ such
that $\Sigma_{\kappa,R}(P_0)<\Sigma_R(\Hbk)$, where
\begin{align*}
\Sigma_{\kappa,R}(P)&=\inf\big\{\la \vphi,H_{\kappa}(P)\vphi\ra\, |\,
 \vphi\in\mathcal{D}_R,\ \|\vphi\|=1\big\},\\
\Sigma_R(\Hbk)&=\inf\big\{\la \vphi,\Hbk\vphi\ra\, |\, \vphi\in\D(\Hbk),
 \|\vphi\|=1 \ \mbox{and}\ \vphi(x_1,x_2)=0\ \\
&\hspace{2cm} \mbox{if}\ |x_1-x_2|<R\big\}.
\end{align*}
Set $\gamma_R=\Sigma_R(\Hbk)-\Sigma_{\kappa,R}(P_0)>0$. There exists a
$\vphi\in\mathcal{D}_R$ so that $\|\vphi\|=1$ and $\la \vphi,
H_{\kappa}(P_0)\vphi\ra\le \Sigma_R(\Hbk)-\gamma_R/2$. Since $\la \vphi,
H_{\kappa}(P)\vphi\ra$ is continuous in $P$, there is a $\delta>0$ such
that, for all $P$ with $|P-P_0|\le\delta$, $\la\vphi,
H_{\kappa}(P)\vphi\ra\le \Sigma_R(\Hbk)-\gamma_R/4$. Choose $f\in
C_0^{\infty}(\BbbR^3)$ as $\mathrm{supp} f\subseteq \{P\in\BbbR^3\, |\,
|P-P_0|\le \delta\}$ with $\|f\|=1$ and define $\vphi_f=f\times \vphi$
for $\vphi\in\mathcal{D}_R$ with $\|\vphi\|=1$. Then we have $\la \vphi_f,
\uni\Hbk\uni^*\vphi_f\ra\le \Sigma_R(\Hbk)-\gamma_R/4$. Notice that
$(\uni^*\vphi_f)(x_1,x_2)=0$ if $|x_1-x_2|<R$. Hence one arrives at $
\Sigma_R(\Hbk)\le \Sigma_R(\Hbk)-\gamma_R/4$ which means a contradiction.

On the other hand, for $\vphi\in \D(\Hbk)$ such
that $\|\vphi\|=1$ and $\vphi(x_1,x_2)=0$ if $|x_1-x_2|<R$, we have
that, by (\ref{FormIonization}),
\begin{align*}
\la \vphi, \Hbk\vphi\ra&=\int\mathrm{d}P\big\la(\uni\vphi)(P),
 H_{\kappa}(P)(\uni\vphi)(P)\big\ra_{L^2(\BbbR^3)\otimes\Fock(L^2(\BbbR^3))}\\
&\ge\int \mathrm{d}P\,
 \tilde{\Sigma}_{R, \kappa}(P)\big\|(\uni\vphi)(P)\big\|^2_{L^2(\BbbR^3)\otimes\Fock(L^2(\BbbR^3))}\\
&\ge\tilde{\Sigma}_{R, \kappa}(0),
\end{align*}
which implies $\Sigma(\Hbk)\ge \Sigma_{\kappa}(0)$ by Lemma \ref{EqSigma} below.
Hence we obtain that  $\Sigma_{\kappa}(0)=\Sigma(\Hbk)$. Finally, we remark
 that, by the slight modification of \cite{Grie}, we can show that
 $\Sigma(\Hbk)=2\Esk$. $\Box$

\begin{lemm}\label{EqSigma}
Let $\tilde{\Sigma}_{R,\kappa}(P)$ be  given by
 (\ref{modifiedThresh}). Then,
\[
 \lim_{R\to \infty}\tilde{\Sigma}_{R,\kappa}(P)=\Sigma_{\kappa}(P).
\]
\end{lemm}
{\it Proof.} It is clear that $\tilde{\Sigma}_{R,\kappa}(P)\le \Sigma_{R,
\kappa}(P)$ which implies $\lim_{R\to\infty}\tilde{\Sigma}_{R,\kappa}(P)\le \Sigma_{
\kappa}(P)$. We will prove the converse. Fix $R$  for a while. For arbitrary $\vepsilon>0$,
there exists $\vphi\in\D(|H_{\kappa}(P)|^{1/2})$ such that  $\|\vphi\|=1$,
$\vphi(\xr)=0$ if $\xr<R$ and  
\[
 \la\vphi, H_{\kappa}(P)\vphi\ra\le
 \tilde{\Sigma}_{R,\kappa}(P)+\frac{\vepsilon}{2}.
\]
For this $\vphi$, there exists a sequence $\{\vphi_n\}\subset
\D(H_{\kappa}(P))$ such that $\|\vphi_n\|=1$, $\lim_{n\to
\infty}\|\vphi-\vphi_n\|=0$ and 
\[
\la\vphi_n, H_{\kappa}(P)\vphi_n\ra\le \la\vphi,
H_{\kappa}(P)\vphi\ra+\frac{\vepsilon}{2}
\]
for all sufficiently large $n$.
Let $\chi$ and $\bar{\chi}$ be the two localization functions with
$\chi^2+\bar{\chi}^2=1$, $\chi$ is identically one on the unit ball
and vanishing  outside the ball of radius $2$. We introduce
$\chi_R(\xr)=\chi(2\xr/R)$ and $\bar{\chi}_R(\xr)=\bar{\chi}(2\xr/R)$.
Then, since $\bar{\chi}_R\vphi_n\in\D(H_{\kappa}(P))$ and
$(\bar{\chi}_R\vphi_n)(\xr)=0$ if $|\xr|<R/2$, we get, by the IMS
localization formula, that 
\begin{align*}
\la\vphi_n, H_{\kappa}(P)\vphi_n\ra=&\la\vphi_n,\chi_R
 H_{\kappa}(P)\chi_R\vphi_n\ra+\la\vphi_n,\bar{\chi}_R
 H_{\kappa}(P)\bar{\chi}_R\vphi_n\ra\\
&-\la\vphi_n, (\nabla_{\xr}\chi_R)^2\vphi_n\ra-\la\vphi_n, (\nabla_{\xr}\bar{\chi_R})^2\vphi_n\ra\\
\ge& E_{\kappa}(P)\|\chi_R\vphi_n\|^2+\Sigma_{R/2,\kappa}(P)\|\bar{\chi}_R\vphi\|^2-\frac{C}{R^2}, 
\end{align*} 
where $C$ is a positive constant independent of $n$. 
Combining these results, we arrive at
\[
 E_{\kappa}(P)\|\chi_R\vphi_n\|^2+\Sigma_{R/2,\kappa}(P)\|\bar{\chi}_R\vphi_n\|^2-\frac{C}{R^2}\le \tilde{\Sigma}_{R,\kappa}(P)+\vepsilon.
\]
First, we take $n\to \infty$. Notice that $\slim\chi_R\vphi_n=0$ and $\slim\bar{\chi}_R\vphi_n=\vphi$. Hence,
\[
 \Sigma_{R/2,\kappa}(P)-\frac{C}{R^2}\le \tilde{\Sigma}_{R,\kappa}(P)+\vepsilon.
\]
Since $\vepsilon$ is arbitrary, we have that $ \Sigma_{R/2,\kappa}(P)-C/R^2\le \tilde{\Sigma}_{R,\kappa}(P)$.
Next, we take $R\to \infty$, then we get the desired result. $\Box$

\subsection{Existence of a  ground state under  the ultraviolet cutoff}
We define  the binding energy with the  ultraviolet cutoff $\kappa$ by 
\[
 E_{\mathrm{bin},\kappa}=2\Esk-\Ebk.
\]
We remark that, by Proposition \ref{LimitEnergy}, $\lim_{\kappa\to\infty}E_{\mathrm{bin},\kappa}=E_{\mathrm{bin}}$.
In this subsection, we will prove the following proposition.

\begin{Prop}\label{UVexistence}
\begin{align}
\inf \mathrm{ess.\, spec}(H_{\kappa}(P))-E_{\kappa}(P)\ge \min\{1,E_{\mathrm{bin},\kappa}\}-\frac{P^2}{4}.
\end{align}
\end{Prop}
\begin{rem}
{\rm Since the dispersion relation for the phonon is constant $1$, we can
 not apply the method developed in \cite{DG1, LLG, LMS} directly. The
 main purpose of this subsection is to show how to overcome this difficulty.}
\end{rem}
Before we enter the proof, we note  the following.
\medskip\\
{\large{\sf Proof of Theorem \ref{GapSpec}}}\\
The assertion directly follows from Proposition \ref{UVexistence}, \ref{LimitEnergy} and
\ref{BottomConv}. $\Box$
\medskip\\\medskip\\

Let $j_1$ and $j_2$ be two smooth localization functions so that
$j_1^2+j_2^2=1$ and $j_1$ is supported in a ball of radius $L$. We
introduce a linear operator $j$ from  $L^2(\BbbR^3)$ to $L^2(\BbbR^3)\oplus
L^2(\BbbR^3)$ by
\[
 jf=j_1(-\im\nabla_k)f\oplus j_2(-\im\nabla_k)f
\]
for $f\in L^2(\BbbR^3)$. Note that $j^*j=\one$. Let $\mathsf{U}$ be the
 unitary operator
 from $\Fock(L^2(\BbbR^3)\oplus L^2(\BbbR^3))$ to
 $\Fock(L^2(\BbbR^3))\otimes\Fock(L^2(\BbbR^3))$ defined by
\begin{align*}
&\mathsf{U}a(f_1\oplus g_1)^*\cdots a(f_n\oplus g_n)^*\Omega\\
=&[a(f_1)^*\otimes\one+\one\otimes
 a(g_1)^*]\cdots[a(f_n)^*\otimes\one+\one\otimes
 a(g_n)^*]\Omega\otimes\Omega,\\
&\mathsf{U}\Omega=\Omega\otimes\Omega.
\end{align*}
We set 
\begin{align*}
\check{\Gamma}(j)=\mathsf{U}\Gamma(j)\, :\, \Fock(L^2(\BbbR^3))\to \Fock(L^2(\BbbR^3))\otimes\Fock(L^2(\BbbR^3)).
\end{align*}
Then $\check{\Gamma}(j)$ is also isometry and we have the following
localization formula in a  similar way to  \cite{LMS}, see also \cite[Lemma A.1]{LLG}.
\begin{lemm}\label{LowerEnergy}
Let $H_{\kappa}^{\otimes}(P)$ be the  self-adjoint operator on
 $L^2(\BbbR^3)\otimes\Fock(L^2(\BbbR^3))\otimes\Fock(L^2(\BbbR^3))$
 defined by
\begin{align*}
&\frac{1}{4}\big(P-\one\otimes\Pf\otimes\one-\one\otimes\one\otimes\Pf\big)^2+\Big(-\Delta_{\xr}+\frac{\alpha
 U_0}{|\xr|}\Big)\otimes\one\otimes\one\\
+&\one\otimes\Nf\otimes\one+\one\otimes\one\otimes\Nf\\
+&
2\sqrt{\alpha}\lambda_0\int_{|k|\le\kappa}\frac{\mathrm{d}k}{(2\pi)^{3/2}|k|}\cos\frac{k\cdot\xr}{2}\otimes[a(k)+a(k)^*]\otimes\one.
\end{align*}
\begin{itemize}
\item[{\rm (i)}] Let $\chi$ be a smooth nonnegative function on
		 $\BbbR^3$ that is compactly supported. Then, for
		 $\vphi\in
		 C_0^{\infty}(\BbbR^3)\hat{\otimes}\Ffin(C_0^{\infty}(\BbbR^3))$,
\begin{align*}
\la\chi\vphi, H_{\kappa}(P)\chi\vphi\ra=\la\check{\Gamma}(j)^*\chi\vphi, H_{\kappa}^{\otimes}(P)\check{\Gamma}(j)\chi\vphi\ra+o_L(\vphi),
\end{align*}
where $o_L(\vphi)$ is the error term which satisfies
\[
 |o_L(\vphi)|\le \tilde{o}(L^0)(\|H_{\kappa}(P)\vphi\|^2+\|\vphi\|^2).
\]
Here $\tilde{o}_L(L^0)$ is a function of $L$ does not depend  on $\vphi$ and
		 vanishes as $L\to \infty$. 
\item[{\rm (ii)}] Let
		 $\Delta_{\kappa}(P)=E_{\kappa}(0)-E_{\kappa}(P)+1$.
For $\vphi\in \D(H^{\otimes}_{\kappa}(P))$,
\[
 \la\vphi, H_{\kappa}^{\otimes}(P)\vphi\ra\ge \la\vphi, [E_{\kappa}(P)+(\one-P_{\Omega})\Delta_{\kappa}(P)]\vphi\ra,
\]
where $P_{\Omega}$ is the orthogonal projection onto $L^2(\BbbR^3)\otimes\Fock(L^2(\BbbR^3))\otimes\Omega$.
\end{itemize}
\end{lemm}

Let $\phi$ and $\bar{\phi}$ be smooth nonnegative functions with
$\phi^2+\bar{\phi}^2=1$, $\phi$ identically one on the unit ball, and
vanishing outside the ball of radius $2$. Define $\phi_R(x)=\phi(x/R)$
and $\bar{\phi}_R(x)=\bar{\phi}(x/R)$. It is not hard to see that, for
$\Psi\in C_0^{\infty}(\BbbR^3)\hat{\otimes}\Ffin(C_0^{\infty}(\BbbR))$,
\begin{align}
\la\Psi, H_{\kappa}(P)\Psi\ra=&\la\phi_R\Psi,
 H_{\kappa}(P)\phi_R\Psi\ra+\la\bar{\phi}_R\Psi, H_{\kappa}(P)\bar{\phi}_R\Psi\ra\nonumber
 \\
&-\la\Psi,(\nabla_{\xr}\phi_R)^2\Psi\ra-\la\Psi,(\nabla_{\xr}\bar{\phi}_R)^2\Psi\ra.\label{IMS}
\end{align}
By Proposition \ref{PropThreshold}, we get
\begin{align}
\la\bar{\phi}_R\Psi, H_{\kappa}(P)\bar{\phi}_R\Psi\ra&\ge
 \Sigma_{\kappa,R}(P)\|\bar{\phi}_R\Psi\|^2\nonumber\\
&\ge \Sigma_{\kappa}(P)\|\bar{\phi}_R\Psi\|^2+\tilde{o}(R^0)\|\Psi\|^2\nonumber\\
&\ge \Sigma_{\kappa}(0)\|\bar{\phi}_R\Psi\|^2+\tilde{o}(R^0)\|\Psi\|^2\nonumber\\
&= 2\Esk\|\bar{\phi}_R\Psi\|^2+\tilde{o}(R^0)\|\Psi\|^2,\label{LowerSigma}
\end{align}
where $\Sigma_{\kappa,R}(P)=\inf_{\vphi\in\mathcal{D}_R,
\|\vphi\|=1}\la\vphi, H_{\kappa}(P)\vphi\ra$.
On the other hand, by Lemma \ref{LowerEnergy} and the fact $\|P_{\Omega}\phi_R\otimes\check{\Gamma}(j)\Psi\|=\|\phi_R\otimes\Gamma(j_1(-\im\nabla_k))\Psi\|$, we obtain
\begin{align}
\la\phi_R\Psi,H_{\kappa}(P)\phi_R\Psi\ra\ge& (E_{\kappa}(P)+\Delta_{\kappa}(P))\|\phi_R\Psi\|^2\nonumber\\
&-\Delta_{\kappa}(P)\|\phi_R\otimes\Gamma(j_1(-\im\nabla_k))\Psi\|^2+\tilde{o}(L^0)\|\Psi\|^2_{H_{\kappa}(P)},\label{LowerLocal}
\end{align}
where $\|\vphi\|_A^2=\|A\vphi\|^2+\|\vphi\|^2$ for a self-adjoint
operator $A$. To summarize, by combining (\ref{IMS}),
(\ref{LowerSigma}),  (\ref{LowerLocal}) and the facts
$\Delta_{\kappa}(P)\ge 1-P^2/4$ and $2\Esk-E_{\kappa}(P)\ge E_{\mathrm{bin},\kappa}-P^2/4$ which follow from Proposition
\ref{PropEnergy} (i), 
 we have the following.

\begin{lemm}\label{SpecInqLemma}
For $\Psi\in\D(H_{\kappa}(P))$,
\begin{align}
\la \Psi, H_{\kappa}(P)\Psi\ra\ge& \Big(E_{\kappa}(P)+\min\{1,E_{\mathrm{bin},\kappa}\}-\frac{P^2}{4}\Big)\|\Psi\|^2\nonumber\\
&-\Delta_{\kappa}(P)\|\phi_R\otimes\Gamma(j_1(-\im\nabla_k))\Psi\|^2+\mathcal{o}(1)\|\Psi\|^2_{H_{\kappa}(P)},\label{SpecInq}
\end{align}
where $o(1)$ is the error term vanishing uniformly in $\Psi$ as both
 $L,R\to \infty$.
\end{lemm}

 We set 
\[
 \BbbR^3_{\le \kappa}=\{k\in\BbbR^3\, |\, |k|\le\kappa\},\ 
\BbbR^3_{> \kappa}=\{k\in\BbbR^3\, |\, |k|>\kappa\}
\]
for each $\kappa>0$. It is well-known that there exists a unitary
operator $V_{\kappa}$ such that 
\begin{align}
V_{\kappa}\Fock(L^2(\BbbR^3))&=\Fock(L^2(\BbbR^3_{\le\kappa}))\otimes\Fock(L^2(\BbbR^3_{>\kappa})),\\
V_{\kappa}a(f)V_{\kappa}^*&=a(f_{\le\kappa})\otimes\one+\one\otimes a(f_{>\kappa})
\end{align}
with $f_{\le\kappa}=\chi_{\kappa}f$ and
$f_{>\kappa}=(1-\chi_{\kappa})f$. (Here $\chi_{\kappa}(k)=1$ for $|k|\le \kappa$,  $\chi_{\kappa}(k)=0$ otherwise.) We also note that, for a
multiplication operator $h$ by the function $h(k)$,
\[
 V_{\kappa}\dG(h)V_{\kappa}^*=\dG(h_{\le \kappa})\otimes\one+\one\otimes\dG(h_{>\kappa}).
\]
In particular,
\begin{align}
V_{\kappa}N_{\mathrm{f}}V_{\kappa}^*=N_{\le
 \kappa}\otimes\one+\one\otimes N_{>\kappa}
\end{align}
where $N_{\le\kappa}$ and $N_{>\kappa}$ are the number operators on
$\Fock(L^2(\BbbR^3_{\le\kappa}))$ and $\Fock(L^2(\BbbR^3_{>\kappa}))$,
respectively. For  notational symplicity, we  denote the unitary operator $\one\otimes
V_{\kappa}$ acting in $L^2(\BbbR^3)\otimes\Fock(L^2(\BbbR^3))$ by the
 symbol $V_{\kappa}$. Let
$\Hil_{\kappa}=L^2(\BbbR^3)\otimes\Fock(L^2(\BbbR^3_{\le \kappa}))$. Then, we can easily see that 
\begin{align}
V_{\kappa}L^2(\BbbR^3)\otimes\Fock(L^2(\BbbR^3))&=\Hil_{\kappa}\otimes
 \Fock(L^2(\BbbR^3_{>\kappa}))\nonumber\\
&=\Hil_{\kappa}\oplus\bigoplus_{n=1}^{\infty}\Big[\Hil_{\kappa}\otimes\big(\otimes_{\mathrm{s}}^nL^2(\BbbR^3_{>\kappa})\big)\Big]\nonumber\\
&=\Hil_{\kappa}\oplus\bigoplus_{n=1}^{\infty}L^2_{\mathrm{sym}}\Big(\underbrace{\BbbR^3_{>\kappa}\times\cdots\times\BbbR^3_{>\kappa}}_n;\Hil_{\kappa}\Big),\label{hardPhDec}
\end{align}
where
$L^2_{\mathrm{sym}}\Big(\underbrace{\BbbR^3_{>\kappa}\times\cdots\times\BbbR^3_{>\kappa}}_n;\Hil_{\kappa}\Big)$
is the $\Hil_{\kappa}$-valued symmetric $L^2$-space on\\
$\underbrace{\BbbR^3_{>\kappa}\times\cdots\times\BbbR^3_{>\kappa}}_n$.
Under the natural identification (\ref{hardPhDec}), the Hamiltonian
$H_{\kappa}(P)$ can be identified as
\begin{align}
&V_{\kappa}H_{\kappa}(P)V_{\kappa}^*\nonumber\\
=&H_{\le
 \kappa}(P)\oplus\bigoplus_{n=1}^{\infty}\Big[\int^{\oplus}_{|k_1|,\dots,|k_n|>\kappa}\Big(H_{\le
 \kappa}\big(P-\sum_{j=1}^nk_j\big)+n\Big)\, \mathrm{d}k_1\cdots\mathrm{d}k_n\Big],\label{HamiIdentification}
\end{align}
where
\begin{align*}
H_{\le\kappa}(P)&=\frac{1}{4}(P-\one\otimes P_{\mathrm{f},\le
 \kappa})^2+\Big(-\Delta_{\xr}+\frac{\alpha U_0}{|\xr|}\Big)\otimes\one+\one\otimes
 N_{\le\kappa}\\
+&2\sqrt{\alpha}\lambda_0\int_{|k|\le \kappa}\frac{\mathrm{d}k}{(2\pi)^{3/2}|k|}\cos\frac{k\cdot\xr}{2}\otimes[a(k)+a(k)^*]
\end{align*}
which is acting in $\Hil_{\kappa}$ and 
$P_{\mathrm{f},\le \kappa}=\int_{|k|\le \kappa}\mathrm{d}k\, ka(k)^*a(k)$.
We note  that, by  the Kato-Rellich theorem, $H_{\le \kappa}(P)$ is
self-adjoint on $\D(-\Delta_{\xr}\otimes\one)\cap \D(\one\otimes
P_{\mathrm{f},\le \kappa})\cap \D(\one\otimes N_{\le \kappa})$ for all
$P$. Therefore, by the closed graph theorem, there exists a positive
constant $C$ such that 
\begin{align}
\|(-\Delta_{\xr}\otimes\one+\one\otimes P_{\mathrm{f},\le
 \kappa}^2+\one\otimes N_{\le \kappa})\vphi\|\le C(\|H_{\le \kappa}(P)\vphi\|+\|\vphi\|)\label{SoftL}
\end{align}
for $\vphi\in\D(-\Delta_{\xr}\otimes\one)\cap \D(\one\otimes
P_{\mathrm{f},\le \kappa}^2)\cap \D(\one\otimes N_{\le \kappa})$.

\begin{lemm}\label{SoftHamiSpec}
Let
 $C_{\kappa}(P)=E_{\kappa}(P)+\min\{1,E_{\mathrm{bin},\kappa}\}-P^2/4$.
\[
 \inf\mathrm{ess.\, spec}(H_{\le \kappa}(P))\ge C_{\kappa}(P).
\]
\end{lemm}
{\it Proof.}  By Lemma \ref{SpecInqLemma}, we get
\begin{align}
&\la\psi, H_{\le \kappa}(P)\psi\ra\nonumber\\
\ge& C_{\kappa}(P)\|\psi\|^2-\Delta_{\kappa}(P)\|\phi_R\otimes\Gamma(j_1(-\im\nabla_k))V_{\kappa}^*\psi\otimes\Omega_{>\kappa}\|^2+o(1)\|\psi\|^2_{H_{\le\kappa}(P)}\label{INQ1}
\end{align}
for $\psi\in\D(H_{\le\kappa}(P))$, where $\Omega_{>\kappa}$ is the Fock
vacuum in $\Fock(L^2(\BbbR^3_{>\kappa}))$.
By Weyl's criterion, for any $\lambda\in\mathrm{ess.\,  spec}(H_{\le \kappa}(P))$,
there is  a normalized sequence $\{\psi_n\}\subset \D(H_{\le \kappa}(P))$
such that  $\wlim\psi_n=0$ and
$\lim_{n\to\infty}\|(H_{\le\kappa}(P)-\lambda)\psi_n\|=0$. Then, by (\ref{INQ1}),
\begin{align}
&\la\psi_n, H_{\le\kappa}(P)\psi_n\ra\nonumber\\
\ge& C_{\kappa}(P)-\Delta_{\kappa}(P)\|\phi_R\otimes\Gamma(j_1(-\im\nabla_k))V_{\kappa}^*\psi_n\otimes\Omega_{>\kappa}\|^2+o(1)\|\psi_n\|^2_{H_{\le\kappa}(P)}.\label{HamiInq2}
\end{align}
We remark that, by (\ref{SoftL}),  
\begin{align*}
&\big\la V_{\kappa}^*\psi_n\otimes\Omega_{>\kappa}, \one\otimes \Nf
 V_{\kappa}^*\psi_n\otimes\Omega_{>\kappa}\big\ra\\
=&\la\psi_n,\one\otimes N_{\le \kappa}\psi_n\ra\le C <\infty,
\end{align*}
where $C$ is a positive constant independent of $n$.
From this, it follows that 
\begin{align}
 \|\phi_R\otimes (\one-\chi_M(\Nf))\Gamma(j_1(-\im\nabla_k))V_{\kappa}^*\psi_n\otimes\Omega_{>\kappa}\|\le \frac{\mbox{Const.}}{M}.
\end{align}
Let $\eta$ be a continuous positive function on $\BbbR^3$ that is
 identically one on the unit ball, and vanishing outside the ball of
 radius 2. Set $\eta_{\kappa}(k)=\eta(k/\kappa)$. We note that 
\begin{align}
V_{\kappa}^*\psi_n\otimes\Omega_{>\kappa}=\one\otimes\Gamma(\eta_{\kappa})V_{\kappa}^*\psi_n\otimes\Omega_{>\kappa}
\end{align}
for all $n\in \BbbN$.
Hence, we obtain
\begin{align*}
&\|\phi_R\otimes\chi_M(\Nf)\Gamma(j_1(-\im\nabla_k))V_{\kappa}^*\psi_n\otimes\Omega\|^2\\
=&\big\la
 (-\Delta_{\xr}+\one)^{1/2}\otimes\chi_M(\Nf)\Gamma(j_1(-\im\nabla_k))V_{\kappa}^*\psi_n\otimes\Omega_>\kappa, \\&\hspace{1cm}(-\Delta_{\xr}+\one)^{-1/2}\phi_R^2\otimes\chi_M(\Nf)\Gamma(j_1(-\im\nabla_k))\Gamma(\eta_{\kappa})V_{\kappa}^*\psi_n\otimes\Omega_{>\kappa}\big\ra.
\end{align*}
It is not hard to check that, by (\ref{SoftL}), 
\begin{align*}
& \|(-\Delta_{\xr}+\one)^{1/2}\otimes\chi_M(\Nf)\Gamma(j_1(-\im\nabla_k))V_{\kappa}^*\psi_n\otimes\Omega_{>\kappa}\|^2\\
\le&\mathrm{Const.}\  \la \psi_n, (H_{\le \kappa}(P)+\one)\psi_n\ra.
\end{align*}
The right hand side of this inequality is uniformly bounded in
$n$. Furthermore,
$(-\Delta_{\xr}+\one)^{-1/2}\phi_R^2\otimes\chi_M(\Nf)\Gamma(j_1(-\im\nabla_k))\Gamma(\eta_{\kappa})$
is a compact operator which implies
\begin{align*}
\slim(-\Delta_{\xr}+\one)^{-1/2}\phi_R^2\otimes\chi_M(\Nf)\Gamma(j_1(-\im\nabla_k))\Gamma(\eta_{\kappa})V_{\kappa}^*\psi_n\otimes\Omega_{>\kappa}=0.
\end{align*}
From these facts, one concludes that 
\[
 \lim_{n\to\infty}\|\phi_R\otimes\Gamma(j_1(-\im\nabla_k))V_{\kappa}^*\psi_n\otimes\Omega_{>\kappa}\|=0
\]
and, by (\ref{HamiInq2}),
\begin{align}
\lambda\ge  C_{\kappa}(P)+o(1)(\lambda^2+1).
\end{align}
Taking $L\to \infty$ and $R\to \infty$, we obtain the desired result. $\Box$
\medskip\\\medskip\\
{\large{\sf  Proof of Proposition \ref{UVexistence}}}\\
By (\ref{HamiIdentification}), we have
\begin{align}
\inf\mathrm{ess.\, spec}(H_{\kappa}(P))=\min\big\{\inf\mathrm{ess.\,
 spec}(H_{\le \kappa}(P)), \tau(P)\big\},
\end{align}
where
\[
 \tau(P)=\inf_{n\ge 1}\inf_{k_1,\dots,k_n\in\BbbR^3_{>\kappa}}\Big[\inf
 \mathrm{spec}\Big(H_{\le \kappa}\big(P-\sum_{j=1}^nk_j\big)\Big)+n\Big].
\]
First, we show that 
\begin{align}
\tau(P)\ge E_{\kappa}(0)+1.\label{TauInq}
\end{align}
Since 
\[
 \la V_{\kappa}^*f\otimes\Omega_{>\kappa},
 H_{\kappa}(P)V_{\kappa}^*f\otimes\Omega_{>\kappa}\ra=\la f, H_{\le \kappa}(P)f\ra,
\]
we have that 
\[
 E_{\kappa}(P)\le \inf \mathrm{spec}(H_{\le \kappa}(P))
\]
for all $P$. Combining this with Proposition \ref{PropEnergy} (iii), we can
easily see (\ref{TauInq}).

From Proposition \ref{PropEnergy} (i), Lemma \ref{SoftHamiSpec} and
(\ref{TauInq}), it follows that 
\begin{align*}
&\inf\mathrm{ess.\, spec}(H_{\kappa}(P))-E_{\kappa}(P)\\
\ge&
 \min\Big\{\min\big\{1,
 E_{\mathrm{bin},\kappa}\big\}-\frac{P^2}{4},
 E_{\kappa}(0)-E_{\kappa}(P)+1\Big\}\\
\ge &\min\Big\{\min\{1, E_{\mathrm{bin},\kappa}\}-\frac{P^2}{4},
 1-\frac{P^2}{4}\Big\}\\
=&\min\big\{1,E_{\mathrm{bin},\kappa}\big\}-\frac{P^2}{4}. \ \ \ \Box
\end{align*}

\appendix

\section{Self-adjointness, fiber decomposition}
\subsection{Proof of Theorem \ref{DefHami} (i)}
The basic idea of the proof is due to Nelson \cite{Nelson}.
Let $K<\kappa$, and let the linear operator $T_{\kappa,K}$ be given by
\begin{align*}
T_{\kappa,K}=\sum_{j=1,2}\int_{|k|\le
 \kappa}\mathrm{d}k\, \beta_K(k)\big[\ex^{\im k\cdot x_j}\otimes
 a(k)-\ex^{-\im k\cdot x_j}\otimes a(k)^*\big]
\end{align*}
with 
\[
 \beta_K(k)=-\frac{\sqrt{\alpha}\lambda_0}{(2\pi)^{3/2}|k|(1+k^2/2)}(1-\chi_K(k)),
\]
where $\chi_K(k)=1$ for $|k|\le K$,  $\chi_K(k)=0$ otherwise.
$T_{\kappa,K}$ is a skew symmetric operator. We denote the closure of
$T_{\kappa,K}$ by the same symbol. Then $T_{\kappa,K}$ is a skew-adjoint
operator: $T_{\kappa,K}^*=-T_{\kappa,K}$. The unitary operator
$U_{\kappa,K}=\ex^{T_{\kappa,K}}$
is called the {\it Gross transformation}. We can easily observe that
\begin{align}
\Uni p_j\otimes\one \Uni^*&=p_j\otimes\one-\A(x_j)-\A(x_j)^*,\\
\Uni \one\otimes a(k)\Uni^*&=\one\otimes a(k)+\sum_{j=1,2}\beta_K(k)\chi_{\kappa}(k)\ex^{-\im
 k\cdot x_j}\otimes\one,
\end{align}
where 
\[
 \A(x)=\int_{|k|\le \kappa}\mathrm{d}k\, k\beta_K(k)\ex^{\im k\cdot
 x}\otimes a(k)
\]
and we use the symbol $p_j=-\im\nabla_{x_j}\, (j=1,2)$.
Using these formulae one gets
\begin{align}
\Uni \Hbk \Uni^*=\Hk
\end{align}
on $C_0^{\infty}(\BbbR^6)\hat{\otimes}\Ffin(L^2(\BbbR^3))$, where 
\begin{align}
\Hk=&\sum_{j=1,2}\Big\{-\frac{1}{2}\Delta_j\otimes \one
 +\frac{1}{2}\Big(-2p_j\cdot \A(x_j)-2\A(x_j)^*\cdot
 p_j\nonumber\\
&+\A(x_j)^2+\A(x_j)^{*2}+2\A(x_j)^*\cdot \A(x_j)\Big)\nonumber\\
&+\sqrt{\alpha}\lambda_0 \int_{|k|\le K}\mathrm{d}k\,
 \frac{1}{(2\pi)^{3/2}|k|}\Big(\ex^{\im k\cdot x_j}\otimes
 a(k)+\ex^{-\im k\cdot x_j}\otimes a(k)^*\Big)\Big\}\nonumber\\
&+\one\otimes\Nf+V_{\kappa,K}(x_1-x_2)\otimes\one+\frac{\alpha U_0}{|x_1-x_2|}\otimes\one+E_{\kappa,K},\label{GrossHami}\\
V_{\kappa,K}(x_1-x_2)=&\sum_{i\neq j}\int_{|k|\le
 \kappa}\mathrm{d}k\Big\{\beta_K(k)^2+\frac{2\sqrt{\alpha}\lambda_0}{(2\pi)^{3/2}|k|}\beta_K(k)\Big\}\ex^{-\im
 k\cdot(x_i-x_j)},\nonumber\\
E_{\kappa,K}=&-2\alpha\lambda_0^2\int_{K\le |k|\le \kappa}\mathrm{d}k\, \frac{1}{(2\pi)^3(1+k^2/2)|k|^2}.\nonumber
\end{align}
Notice that $E_{\kappa,K}$   is finite even for  $\kappa=\infty$. $\Hk$ is
closable and we denote its closure by the same symbol.

\begin{Prop}\label{SAUNI}
For any $\alpha<\infty, U_0<\infty, \kappa<\infty$ and $K$, $\Hk$ is self-adjoint on $\D(L_{\mathrm{bp}})$, essentially
 self-adjoint on any core for  $L_{\mathrm{bp}}$ and bounded from below. Moreover
\[
 \Uni \Hbk\Uni^*=\Hk.
\]
\end{Prop}
{\it Proof.} By the inequality (\ref{NumberEst}), and
\[
 \|a(f)^{\#}a(g)^{\#}\vphi\|\le 8\|f\| \|g\|\|(\Nf+\one)\vphi\|,
\]
one can check that 
\[
 \|\Hk\vphi\|\le C(\|L_{\mathrm{bp}}\vphi\|+\|\vphi\|),\ \vphi\in\D(L_{\mathrm{bp}})
\]
with some positive constant $C<\infty$. (Note  that the finiteness
of  $\kappa$ is crucial  here.) From this we have
\begin{align}
 \|\Hbk\Uni\vphi\|=\|\Uni^*\Hbk\Uni\vphi\|\le C(\|L_{\mathrm{bp}}\vphi\|+\|\vphi\|)\label{Ineq1}
\end{align}
for $\vphi\in C_0^{\infty}(\BbbR^6)\widehat{\otimes}\Ffin(C_0^{\infty}(\BbbR^3))$. Since
$\D(\Hbk)=\D(L_{\mathrm{bp}})$, we have
\[
 \|L_{\mathrm{bp}}\Uni\vphi\|\le
 C^{'}(\|L_{\mathrm{bp}}\vphi\|+\|\vphi\|),\ \vphi\in
 C_0^{\infty}(\BbbR^6)\widehat{\otimes}\Ffin(C_0^{\infty}(\BbbR^3))
\]
by the closed graph theorem and (\ref{Ineq1}).  Thus we conclude that
$\Uni\D(L_{\mathrm{bp}})\subseteq \D(L_{\mathrm{bp}})$. Similarly
$\Uni^*\D(L_{\mathrm{bp}})\subseteq \D(L_{\mathrm{bp}})$ and hence
$\D(\Uni
\Hbk\Uni^*)\\=\D(\Uni L_{\mathrm{bp}}\Uni^*)=\D(L_{\mathrm{bp}})=\D(\Hbk)$.
Since
\[
 \Uni\Hbk\Uni^*\vphi=\Hk\vphi
\]
for all $\vphi\in C_0^{\infty}(\BbbR^6)\widehat{\otimes}\Ffin(C_0^{\infty}(\BbbR^3))$, we conclude that
$\Uni^*\Hbk\Uni=\Hk$ as an operator equality. $\Box$

\hspace{0.5cm}

The quadratic  form
\begin{align}
&B_{\kappa,K}(\vphi,\psi)\nonumber\\
=&\sum_{j=1,2}\Big\{-\la p_j\otimes\one\vphi,
 \A(x_j)\psi\ra-\la \A(x_j)\vphi, p_j\otimes\one\psi\ra\nonumber\\
&+\frac{1}{2}\la \vphi,
 \A(x_j)^2\psi\ra+\frac{1}{2}\la \A(x_j)^2\vphi, \psi\ra
+\la A_{\kappa,K}(x_j)\vphi, A_{\kappa,K}(x_j)\psi\ra\Big\}\nonumber\\
&+\la \vphi, H_{IK}\psi\ra
+\la \vphi, V_{\kappa,K}(x_1-x_2)\otimes \one \psi\ra+\la \vphi,
 \frac{\alpha U_0}{|x_1-x_2|}\otimes \one\psi\ra+E_{\kappa,K}\la\vphi,\psi\ra\label{BForm}
\end{align}
is well defined on $\D(L_{\mathrm{bp}}^{1/2})\times
\D(L_{\mathrm{bp}}^{1/2})$ for all $\kappa\le \infty$ and $K$, where
\[
 H_{IK}=\sqrt{\alpha}\lambda_0\sum_{\j=1,2} \int_{|k|\le K}
 \frac{\mathrm{d}k}{(2\pi)^{3/2}|k|}\Big[\ex^{\im k\cdot x_j}\otimes
 a(k)+\ex^{-\im k\cdot x_j}\otimes a(k)^*\Big].
\]

\begin{lemm}\label{formEST}
For all $\vepsilon>0$, there is a $0<C_{\vepsilon,K}<\infty$ such that 
\begin{align}
 |B_{\kappa,K}(\vphi,\vphi)|\le (4C(K)^2+4C(K)+\vepsilon)\|L_{\mathrm{bp}}^{1/2}\vphi\|^2+C_{\vepsilon,K}\|\vphi\|^2\label{BEst}
\end{align}
for all $\kappa\le \infty$,
where
\[
 C(K)^2=\int\mathrm{d}k\, k^2\beta_K(k)^2=\int_{|k|>K}\mathrm{d}k\, \frac{\alpha\lambda_0^2}{(2\pi)^3(1+k^2/2)^2}.
\]
\end{lemm}
{\it Proof.} First we note that, for $\vphi\in \D(L_{\mathrm{bp}})$,
\begin{align}
\|p_j\otimes\one\vphi\|&\le \|(L_{\mathrm{bp}}+\one)^{1/2}\vphi\|, \label{PEst}\\
\|A_{\kappa,K}(x_j)^{\#}\vphi\|& \le C(K) \|(L_{\mathrm{bp}}+\one)^{1/2}\vphi\| \label{AniEst}
\end{align}
by (\ref{NumberEst}). From these inequalities, it follows that 
\begin{align*}
|\la p_j\vphi, A_{\kappa,K}(x_j)\vphi\ra|&\le C(K)
 \|(L_{\mathrm{bp}}+\one)^{1/2}\vphi\|^2,\\
|\la\vphi, A_{\kappa,K}(x_j)^2\vphi\ra|&\le  C(K)^2
 \|(L_{\mathrm{bp}}+\one)^{1/2}\vphi\|^2.
\end{align*}
On the other hand, for any $\vepsilon_1>0$, we have
\[
 |\la\vphi, H_{IK}\vphi\ra|\le
  \vepsilon_1\|(L_{\mathrm{bp}}+\one)^{1/2}\vphi\|^2+\frac{4}{\vepsilon_1} C_2(K)\|\vphi\|^2
\]
by (\ref{NumberEst}), where $C_2(K)=\alpha\lambda_0^2\int_{|k|\le K}\mathrm{d}k/(2\pi)^3|k|^2$.
Moreover,
\begin{align*}
 |\la\vphi, V_{\kappa,K}(x_1-x_2)\otimes\one\vphi\ra|&\le 2\int\mathrm{d}k\,
  \Big\{\beta_K(k)^2+\frac{2\sqrt{\alpha}\lambda_0}{(2\pi)^{3/2}|k|}|\beta_K(k)|\Big\}
  \|\vphi\|^2\\
&=:2C_3(K)\|\vphi\|^2
\end{align*}
and, for any $\vepsilon_2>0$, there exists $b_{\vepsilon_2}>0$ such that 
\[
 |\la \vphi, \frac{U_0\alpha}{|x_1-x_2|}\otimes\one\vphi\ra|\le \vepsilon_2\|L_{\mathrm{bp}}^{1/2}\vphi\|^2+b_{\vepsilon_2}\|\vphi\|^2.
\] 
Combining these results, we obtain the desired assertion. $\Box$

\hspace{0.5cm}

Choose $K$ sufficiently large as $4C(K)^2+4C(K)<1$. Then, by Lemma
\ref{formEST} and the KLMN theorem (see, e.g., \cite{ReSi1}), for
$\kappa\le \infty$, there exists a unique self-adjoint operator $H_{\kappa,K}^{\mathrm{bp}'}$
such that  
\[
 \la \vphi, H_{\kappa,K}^{\mathrm{bp}'}\vphi\ra=\la L_{\mathrm{bp}}^{1/2}\vphi, L_{\mathrm{bp}}^{1/2}\vphi\ra+B_{\kappa,K}(\vphi,\vphi).
\]
For $\kappa<\infty$, by Proposition \ref{SAUNI}, we have
\[
 H_{\kappa,K}^{\mathrm{bp}'}=\Hk=\Uni \Hbk \Uni^*.
\]
From this fact, it is natural to denote $H_{\infty,K}^{\mathrm{bp}'}$ as $H_{\infty,K}^{\mathrm{bp}}$.
\begin{lemm}\label{UniBound1}
\[
 \lim_{\kappa\to\infty}B_{\kappa,K}(\vphi,\vphi)=B_{\infty,K}(\vphi,\vphi)
\]
unifromly on any set of $\vphi$ in $\D(L_{\mathrm{bp}}^{1/2})$ for which
 $\|L_{\mathrm{bp}}^{1/2}\vphi\|+\|\vphi\|$ is bounded.
\end{lemm}
{\it Proof. } By the similar argument in the proof of Lemma
\ref{formEST}, we have
\begin{align}
&|B_{\kappa,K}(\vphi,\vphi)-B_{\infty,K}(\vphi,\vphi)|\nonumber\\
\le&
 4(C(\kappa)+2C(K)C(\kappa))\|(L_{\mathrm{bp}}+\one)^{1/2}\vphi\|^2+\big(2C_3(\kappa)+|E_{\infty,K}-E_{\kappa, K}|\big)\|\vphi\|^2,\label{UniBEst} 
\end{align}
where $C(\kappa)$ (resp. $C_3(\kappa)$) is $C(K)$ (resp. $C_3(K)$) with
$K$ replaced by $\kappa$. $\Box$

\hspace{0.5cm}

Applying \cite[Theorem VIII. 25]{ReSi1}, we immediately obtain the
following.
\begin{Prop}\label{NormRes}
For $K$ satisfying $4C(K)^2+4C(K)<1$, $\Hk$ converges to
 $H_{\infty,K}^{\mathrm{bp}}$ as $\kappa\to \infty$ in the norm resolvent sense.
\end{Prop}
{\sf {\large Proof of Theorem \ref{DefHami} (i).}}\\
Since $U_{\kappa,K}$ converges to $U_{\infty, K}$ strongly, we
have the desired assertion by Proposition \ref{NormRes}. $\Box$

\subsection{Proof of Theorem \ref{DefHami} (ii) and (iii)}
Let $\Hk$ be the Hamiltonian given by (\ref{GrossHami}). It is not hard
to see that $\uni\Hk \uni^*$ is also decomposable and 
\[
 \uni\Hk
 \uni^*=\int^{\oplus}_{\BbbR^3}\mathscr{H}_{\kappa,K}^{\mathrm{bp}}(P)\, \mathrm{d}P.
\]
On $C_0^{\infty}(\BbbR^3)\hat{\otimes}\Ffin(C_0^{\infty}(\BbbR^3))$, we
can represent $\mathscr{H}_{\kappa,K}^{\mathrm{bp}}(P)$ as follows,
\begin{align}
&\mathscr{H}_{\kappa, K}^{\mathrm{bp}}(P)\nonumber\\
=&\frac{1}{4}(P-\one\otimes
 \Pf)^2-\Delta_{\xr}\otimes\one\nonumber+\frac{\alpha U_0}{|\xr|}\otimes\one+\one\otimes\Nf\\
&+\sum_{j=1,2}\Big\{-\Big[(-1)^{j-1}(-\im\nabla_{\xr})\otimes\one+\frac{1}{2}\big(P-\one\otimes\Pf\big)\Big]\cdot
 A_{\kappa,K}\Big((-1)^{j-1}\frac{\xr}{2}\Big)\nonumber\\
&-
 A_{\kappa,K}\Big((-1)^{j-1}\frac{\xr}{2}\Big)^*\cdot\Big[(-1)^{j-1}(-\im\nabla_{\xr})\otimes\one+\frac{1}{2}\big(P-\one\otimes\Pf\big)\Big]\nonumber\\
&+ \frac{1}{2}A_{\kappa,K}\Big((-1)^{j-1}\frac{\xr}{2}\Big)^2+
 \frac{1}{2}A_{\kappa,K}\Big((-1)^{j-1}\frac{\xr}{2}\Big)^{*2}\nonumber\\
&+A_{\kappa,K}\Big((-1)^{j-1}\frac{\xr}{2}\Big)^*\cdot A_{\kappa,K}\Big((-1)^{j-1}\frac{\xr}{2}\Big)\Big\}\nonumber\\
&+2\sqrt{\alpha}\lambda_0\int_{|k|\le K}\frac{\mathrm{d}k}{(2\pi)^{3/2}|k|}\cos\frac{k\cdot\xr}{2}\otimes[a(k)+a(k)^*]\nonumber\\
&+V_{\kappa,K}(\xr)\otimes\one+E_{\kappa,K}\label{GrossFiberHami}
\end{align}
The symmetric operator $\Hk(P)$ is now defined by the right hand side
of (\ref{GrossFiberHami}). Clearly this operator  is closable and we denote its
closure by the same symbol.

\begin{Prop}
For all $\kappa<\infty$, $K<\infty$, $\alpha<\infty$ and $P\in\BbbR^3$,
 $\Hk(P)$ is self-adjoint on $\D(-\Delta_{\xr}\otimes\one)\cap
 \D(\one\otimes\Pf^2)\cap\D(\one\otimes\Nf)$, essentially self-adjoint
 on any core for the self-adjoint operator $L$ defined by (\ref{LOp}).
Moreover,
\begin{align}
\uni\Hk \uni^*=\int^{\oplus}_{\BbbR^3}\Hk(P)\, \mathrm{d}P.\label{DecomH}
\end{align}
\end{Prop}
{\it Proof.}
In the proof of Propsotion \ref{SAUNI}, we have proved that
$\D(\Uni L_{\mathrm{bp}}\Uni^*)=\D(L_{\mathrm{bp}})$. Thus, by the closed
graph theorem,  
there is  a constant $C$ such that
\[
 \|\Uni L_{\mathrm{bp}}\Uni^*\vphi\|^2+\|\vphi\|^2\le C(\|L_{\mathrm{bp}}\vphi\|^2+\|\vphi\|^2)
\] 
for all $\vphi\in \D(L_{\mathrm{bp}})$.
Choose $\vphi$ as $\uni\vphi=\eta_n\otimes\psi$ with
$\psi\in C_0^{\infty}(\BbbR^3)\hat{\otimes}\Ffin(C_0^{\infty}(\BbbR^3))$
and 
\begin{align}
\eta_n=n^{3/2}\chi_{M_n(P)},\label{AppVector}
\end{align}
with 
$
M_n(P)=\Big\{k\in\BbbR^3\, \big|\, |k_j-P_j|\le \frac{1}{2n},\,
j=1,2,3\Big\}, 
$
where $\chi_S$ is the characteristic function for the set $S$. 
Then, we get that 
\begin{align*}
\int_{\BbbR^3}\mathrm{d}k\, \eta_n(k)^2\|W_{\kappa,K}L(k)W_{\kappa,K}^*\psi\|^2
\le C\Big(\int_{\BbbR^3}\mathrm{d}k\, \eta_n(k)^2\|L(k)\psi\|^2+\|\psi\|^2\Big),
\end{align*}
where 
\[
 L(P)=\frac{1}{4}(P-\one\otimes\Pf)^2-\Delta_{\xr}\otimes\one+\one\otimes\Nf
\]
and
\[
 W_{\kappa,K}=\exp\Big\{\sum_{j=1,2}\int\mathrm{d}k\,
 \beta_K(k)\Big[\ex^{\im k\cdot (-1)^{j-1}\xr/2}\otimes a(k)-\ex^{-\im k\cdot (-1)^{j-1}\xr/2}\otimes a(k^*)\Big]\Big\}.
\]
Note  here that  we have  used  the following facts:
\begin{align}
\uni U_{\kappa,K}\uni^*&=\int_{\BbbR^3}^{\oplus}W_{\kappa,K}\,
 \mathrm{d}P,\label{UniUni}\\
\uni L_{\mathrm{bp}}\uni^*&=\int_{\BbbR^3}^{\oplus}L(P)\, \mathrm{d}P.\nonumber
\end{align}
 Taking the limit $n\to \infty$, we get
\[
 \|W_{\kappa,K}^*L(P)W_{\kappa,K}\psi\|^2+\|\psi\|^2\le C(\|L(P)\psi\|^2+\|\psi\|^2).
\]
Since $C_0^{\infty}(\BbbR^3)\hat{\otimes}\Ffin(C_0^{\infty}(\BbbR^3))$
is a core for $L(P)$, we can extend this inequailty to
$\D(L(P))=\D(-\Delta_{\xr}\otimes\one)\cap\D(\one\otimes\Pf^2)\cap\D(\one\otimes\Nf)$. 
Thus, we have $W_{\kappa,K}\D(L(P))\subseteq \D(L(P))$ for all
$P$. Similarly $W_{\kappa,K}^*\D(L(P))\subseteq \D(L(P))$ and we
conclude that 
\[
 \D(W_{\kappa,K}H_{\kappa}(P)W_{\kappa,
 K}^*)=\D(W_{\kappa,K}L(P)W_{\kappa, K}^*)=\D(L(P)).
\]
Since 
\begin{align}
 W_{\kappa, K}H_{\kappa}(P)W_{\kappa,K}^*=\Hk(P)\label{UniHami}
\end{align}
on $C_0^{\infty}(\BbbR^3)\hat{\otimes}\Ffin(C_0^{\infty}(\BbbR^3))$,
we arrive at $W_{\kappa, K}H_{\kappa}(P)W_{\kappa,K}^*=\Hk(P)$ as an
operator equality. Thus,  $\Hk(P)$ is self-adjoint on $\D(L(P))$. To show 
(\ref{DecomH}) is an easy execise. $\Box$

\begin{lemm}\label{AEDecomposable}
$\uni \Hki \uni^*$ is denomposable and can be represented as 
\[
 \uni\Hki
 \uni^*=\int^{\oplus}_{\BbbR^3}\tilde{H}_{\infty,K}^{\mathrm{bp}}(P)\, \mathrm{d}P.
\]
Moreover, for a.e. $P$, $\Hk(P)$ converges to
 $\tilde{H}_{\infty,K}^{\mathrm{bp}}(P)$ in the norm resolvent sense as
 $\kappa\to\infty$.
\end{lemm}
This is a direct consequence of the following abstract theory.
\begin{lemm}
Let $A_n\, (n\in \BbbN)$ and $A$ be self-adjoint operators on a Hilbert space
 $\int^{\oplus}_M\h\, \mathrm{d}\mu(m)$. Suppose that $A_n$ is
 decomposable for all $n\in\BbbN$, i.e., $A_n=\\\int^{\oplus}_MA_n(m)\,
 \mathrm{d}\mu(m)$. Suppose that $A_n$ converges to $A$ in the norm
 resolvent sense as $n\to \infty$. Then,
\begin{itemize}
\item[{\rm (i)}] $A$ is also decomposable. Hence we can represent $A$ as
		 the fiber direct integral $A=\int^{\oplus}_M A(m)\, \mathrm{d}\mu(m)$,
\item[{\rm (ii)}] For $\mu$-a.e. $m$, $A_n(m)$ converges to $A(m)$ in
		 the norm  resolvent sense as $n\to \infty$.
\end{itemize}
\end{lemm}
{\it Proof.} (i) 
$A_n$ is decomposable if and only if $\ex^{\im t A_n}F=F\ex^{\im t
A_n}$ for all $t\in\BbbR$ and $F\in L^{\infty}(M,\mathrm{d}\mu) $. Taking $n\to\infty$, we arrive at 
$\ex^{\im t A}F=F\ex^{\im t A}$ which means that $A$ is decomposable and
can be written as $A=\int^{\oplus}_MA(m)\, \mathrm{d}\mu(m)$.

(ii) For $\mu$-a.e. $m$, we obtain that 
\begin{align*}
\|(A_n(m)+\im)^{-1}-(A(m)+\im)^{-1}\|\le 
\|(A_n+\im)^{-1}-(A+\im)^{-1}\|\ \ \ (n\to \infty). \ \ \ \ \Box
\end{align*}

\vspace{0.5cm}
We note that Lemma \ref{AEDecomposable} guarantees the existence of the
limiting Hamiltonian $\tilde{H}_{\infty,K}^{\mathrm{bp}}(P)$ only for
a.e. $P$. To prove the existence of the limiting Hamiltonian for {\it
all} $P$, we need more technical preparations.

Let $\tilde{B}_{\kappa,K}^P(\vphi, \psi)$ be the quadratic form on
$\D(L(P)^{1/2})\times \D(L(P)^{1/2})$ defined by
\begin{align}
&\tilde{B}_{\kappa,K}^P(\vphi, \psi)\nonumber\\
=&\sum_{j=1,2}\Big\{-\Big\la\Big[(-1)^{j-1}(-\im\nabla_{\xr})\otimes\one+\frac{1}{2}\big(P-\one\otimes\Pf\big)\Big]\vphi,
 A_{\kappa,K}\Big((-1)^{j-1}\frac{\xr}{2}\Big)\psi\Big\ra\nonumber\\
&-
 \Big\la A_{\kappa,K}\Big((-1)^{j-1}\frac{\xr}{2}\Big)\vphi, \Big[(-1)^{j-1}(-\im\nabla_{\xr})\otimes\one+\frac{1}{2}\big(P-\one\otimes\Pf\big)\Big]\psi\Big\ra\nonumber\\
&+ \frac{1}{2}\Big\la \vphi, A_{\kappa,K}\Big((-1)^{j-1}\frac{\xr}{2}\Big)^2\psi\Big\ra+
 \frac{1}{2}\Big\la
 A_{\kappa,K}\Big((-1)^{j-1}\frac{\xr}{2}\Big)^{2}\vphi, \psi\Big\ra\nonumber\\
&+\Big\la A_{\kappa,K}\Big((-1)^{j-1}\frac{\xr}{2}\Big)\vphi,  A_{\kappa,K}\Big((-1)^{j-1}\frac{\xr}{2}\Big)\psi\Big\ra\Big\}\nonumber\\
&+\Big\la\vphi, 2\sqrt{\alpha}\lambda_0\int_{|k|\le K}\frac{\mathrm{d}k}{(2\pi)^{3/2}|k|}\cos\frac{k\cdot\xr}{2}\otimes[a(k)+a(k)^*]\psi\Big\ra\nonumber\\
&+\la\vphi, V_{\kappa,K}(\xr)\otimes\one\psi\ra+\la\vphi,\frac{\alpha U_0}{|\xr|}\otimes\one\psi\ra+E_{\kappa,K}\la\vphi,\psi\ra\label{FormB}
\end{align}
for $K<\kappa\le\infty$.

\begin{lemm}\label{FiberBEst}
\begin{itemize}
\item[{\rm (i)}] For all $\vepsilon>0$, there is a $C_{\vepsilon,K}>0$
		 such that 
\begin{align*}
|\tilde{B}_{\kappa,K}^P(\vphi,\vphi)|\le
 (4C(K)^2+4C(K)+\vepsilon)\|L(P)^{1/2}\vphi\|^2+C_{\vepsilon,K}\|\vphi\|^2.
\end{align*}
\item[{\rm (ii)}]
\begin{align*}
\lim_{\kappa\to\infty}\tilde{B}_{\kappa,K}^P(\vphi,\vphi)=\tilde{B}_{\infty,K}^P(\vphi,\vphi)
\end{align*}
uniformly on any set of $\vphi$ in $\D(L(P)^{1/2})$ for which $\|L(P)^{1/2}\vphi\|^2+\|\vphi\|^2$
is bounded.
\end{itemize}
\end{lemm}
{\it Proof.} (i) Let $\eta_n$ be the vector defined by
(\ref{AppVector}).
Choose $\vphi$ as $\uni\vphi=\eta_n\otimes\psi$ with
$\psi\in\D(L(0)^{1/2})$. Then we have
\begin{align*}
B_{\kappa,K}(\vphi,\vphi)=\int_{\BbbR^3}\mathrm{d}P\, \eta_n(P)^2\tilde{B}_{\kappa,K}^P(\psi,\psi)
\end{align*}
where $B_{\kappa,K}$ is the quadratic form given by (\ref{BForm}).
By Lemma \ref{formEST}, we get
\begin{align*}
&\Big|\int\mathrm{d}P\, \eta_n(P)^2
 \tilde{B}_{\kappa,K}^P(\psi,\psi)\Big|\\
\le&(4C(K)^2+4C(K)+\vepsilon)\int\mathrm{d}P\, \eta_n(P)^2\|L(P)^{1/2}\psi\|^2+C_{\vepsilon,K}\|\psi\|^2.
\end{align*}
Taking the limit $n\to \infty$, we conclude (i). (Here we use the fact
$\D(L(0)^{1/2})=\D(L(P)^{1/2})$ for all $P$.)
Similarly we can prove 
\begin{align}
&|\tilde{B}_{\kappa,K}^P(\psi,\psi)-\tilde{B}_{\infty,K}^P(\psi,\psi)|\nonumber\\
\le&
 4(C(\kappa)+2C(K)C(\kappa))\|(L(P)+\one)^{1/2}\psi\|^2+\big(2C_3(\kappa)+|E_{\infty, K}-E_{\kappa, K}|\big)\|\psi\|^2\label{BTInq}
\end{align}
 by (\ref{UniBEst}). $\Box$
\medskip\\\medskip\\
{{\large \sf Proof of Theorem \ref{DefHami} (ii) and (iii)}}\\
From Lemma \ref{FiberBEst} and the KLMN theorem \cite{ReSi2}, it follows
that, for sufficiently large $K$ as $4C(K)^2+4C(K)<1$, there exists a
unique self-adjoint operator $H_{\kappa, K}^{\mathrm{bp}'}(P)$ such that 
\[
 \la\vphi, H_{\kappa, K}^{\mathrm{bp}'}(P)\vphi\ra=\la L(P)^{1/2}\vphi,
 L(P)^{1/2}\vphi\ra +\tilde{B}_{\kappa,K}^P(\vphi,\vphi).
\] 
For $\kappa<\infty$, it can be easily shown that $H_{\kappa,
K}^{\mathrm{bp}'}(P)=H_{\kappa, K}^{\mathrm{bp}}(P)$. (From now on, we
also denote $H_{\infty,K}^{\mathrm{bp}'}(P)$ by $H_{\infty,K}^{\mathrm{bp}}(P)$.)
 Moreover, by Lemma
\ref{FiberBEst}, $H_{\kappa, K}^{\mathrm{bp}}(P)$ converges to
$H_{\infty, K}^{\mathrm{bp}}(P)$ in the norm resolvent sense for all $P$.
Since $W_{\kappa, K}^*$ converges to $W_{\infty, K}^*$ strongly, we conclude
(ii) by (\ref{UniHami})

Finally we show (iii) in Theorem \ref{DefHami}.
Since $\tilde{H}_{\infty,K}^{\mathrm{bp}}(P)=H_{\infty,
K}^{\mathrm{bp}}(P)$ for a.e. $P$, we have that 
\[
 \int^{\oplus}_{\BbbR^3}\tilde{H}_{\infty,K}^{\mathrm{bp}}(P)\,
 \mathrm{d}P
=\int^{\oplus}_{\BbbR^3}H_{\infty,K}^{\mathrm{bp}}(P)\, \mathrm{d}P. 
\]
Noting that the operator equality (\ref{UniUni}) is valid for $\kappa=\infty$,
 we have the desired assertion. $\Box$

\section{Convergence of the ground state energies and the bottom of
 the essential spectrum}
Let $\Ebk$ and $\Esk$ be the ground state energy for $\Hbk$ and $\Hsk$
 respectively. Further we denote $\inf \mathrm{spec}(H_{\kappa}(P))$,
resp. $\inf\mathrm{spec}(H(P))$,  by $E_{\kappa}(P)$, resp. $E(P)$.
\begin{Prop}\label{LimitEnergy}
For all $\alpha, U_0>0$, the following holds.
\begin{itemize}
\item[{\rm (i)}]$\displaystyle
		\lim_{\kappa\to\infty}\Ebk=\Eb.$
\item[{\rm (ii)}]$\displaystyle \lim_{\kappa\to\infty}\Esk=\Es$.
\item[{\rm (iii)}] $\displaystyle \lim_{\kappa\to
		\infty}E_{\kappa}(P)=E(P)$  for all $P$.
\end{itemize}
\end{Prop}
{\it Proof.} (i) and (iii) are direct consequences of Lemma \ref{UniBound1}  and \ref{FiberBEst}. 
(Note that $\Ebk=\inf\mathrm{spec}(\SHk)$ and
$\Eb=\inf\mathrm{spec}(H_{\infty, K}^{\mathrm{bp}})$. Also note that
$E_{\kappa}(P)=\inf \mathrm{spec}(\SHk(P))$ and
$E(P)=\inf\mathrm{spec}(H_{\infty, K}^{\mathrm{bp}}(P))$ for all $P$.)
We can  show (ii) in a similar way.  $\Box$

\begin{Prop}\label{BottomConv}
For all  $\alpha, U_0>0$,
\begin{align}
\lim_{\kappa\to \infty}\inf\mathrm{ess.\,
 spec}(H_{\kappa}(P))=\inf\mathrm{ess.\, spec}(H(P)).\label{LimitInfEss}
\end{align}
\end{Prop}
{\it Proof.}  Let $\SHk(P)$ be the Hamiltonian defined by the form sum
$L(P)+\tilde{B}_{\kappa,K}^P$ for a sufficiently large $K$, see (\ref{FormB}).
Notice that (\ref{LimitInfEss}) is equivalent to 
\begin{align}
\lim_{\kappa\to \infty}\inf\mathrm{ess.\,
 spec}(\SHk(P))=\inf\mathrm{ess.\, spec}(\Hki(P))\label{LimitInfEss2}
\end{align}
because $W_{\kappa,K}H_{\kappa}(P)W_{\kappa,K}^*=\SHk(P)$ for all
$\kappa\le \infty$.
By Lemma \ref{FiberBEst} (i), we have that, for all $\kappa\le \infty$ and large $K$,
\[
 L(P)+\one\le C(\SHk(P)+\one)
\]
where $C$ is independent of $\kappa$. Combining this with
(\ref{BTInq}), we can conclude that 
\[
 \SHk(P)\le (1+D(\kappa))\Hki(P)+D(\kappa)
\]
and 
\[
 \Hki(P)\le (1+D(\kappa))\SHk(P)+D(\kappa),
\]
where $D(\kappa)$ is a positive constant satisfying $\lim_{\kappa\to
\infty}D(\kappa)=0$.
By the min-max principle, we have that 
\[
 \inf\mathrm{ess.\, spec}(\SHk(P))\le (1+D(\kappa))\inf\mathrm{ess.\, spec}(\Hki(P))+D(\kappa)
\]
and 
\[
 \inf\mathrm{ess.\, spec}(\Hki(P))\le (1+D(\kappa))\inf\mathrm{ess.\, spec}(\SHk(P))+D(\kappa).
\]
Taking the limit $\kappa\to \infty$, we obtain the desired assertion 
(\ref{LimitInfEss2}). $\Box$

\end{document}